\documentclass[twocolumn,showpacs]{revtex4}
\usepackage{graphicx}
\usepackage{dcolumn}
\usepackage{bm}
\usepackage{longtable}
\usepackage{epsfig}
\usepackage{amsmath}

\newcommand{\ima}{{\mbox{Im}\,}}
\newcommand{\rea}{{\mbox{Re}\,}}

\begin{document}

\title{Chiral extrapolation of pion-pion scattering phase shifts \\
within standard and unitarized Chiral Perturbation Theory}

\author{J. Nebreda, J. R. Pel\'aez and G. R\'ios}

\address{Departamento de F\'{\i}sica Te\'orica II. Universidad Complutense, 28040, Madrid. Spain.}

\begin{abstract}
We calculate the  pion-pion elastic scattering phase shifts
for pion masses from the chiral limit to values of interest for
lattice studies. At low energies, we use the standard Chiral Perturbation
Theory expressions to one and two loops. In addition, we study
the phase shifts mass dependence in the resonance region by means of 
dispersion theory
in the form of unitarized Chiral Perturbation Theory and the Inverse 
Amplitude Method.
We pay particular attention to the case when resonances 
are close to threshold, illustrating the different behavior between
scalar and vector resonances. We also provide
the estimation of uncertainties, 
which are dominated by
those of the $O(p^6)$ chiral parameters.
\end{abstract}

\maketitle

\section{Introduction.}
Elastic pion-pion scattering has been an object of study 
for many decades due to several reasons. In particular, 
pions are very relevant in the description of final states in other hadronic processes.
Also, the two pion correlated exchange in the scalar-isoscalar channel
is the main contribution to nucleon-nucleon attraction, 
and has been interpreted
for long as a  scalar ``sigma'' 
resonance \cite{Johnson:1955zz}, whose existence, mass and width have been the subject 
of an intense debate. 
Actually this resonance, nowadays called $f_0(600)$,
appears as a pole deep in the
second Riemann sheet of the scattering amplitude
(see the ``Note on scalar mesons'' 
in \cite{PDG2010} for a detailed
account).
Finally, the pion-pion interaction at low energies is also relevant for the determination of light quark mass ratios and the size of the chiral condensate
\cite{Colangelo:2001sp}.

On the theory side, unfortunately, 
neither the elastic resonance region nor the low energy
region are accessible to perturbative QCD calculations. In order to describe these processes
in terms of quarks and gluons one should rely on lattice techniques.
For a long time, these techniques have found little applications in this low-energy
realm due to complications on the implementation
of chiral symmetry,
the small physical values of the light quarks and other technicalities
as the existence of quark-line disconnected diagrams in some channels.
However, very recently, lattice results have become available
for the 
$\rho(770)$ and $f_0(600)$ resonance masses
\cite{lattice1,lattice2,lattice3,lattice4,lattice5}, 
the pion decay constant \cite{lattice2,fpilattJLQCD,Beane:2007xs} or even the isospin 2
scattering length \cite{Beane:2007xs,Feng:2009ij}, obtained with pion masses which are not too far from 
the physical values. Recent developments \cite{Bulava:2009ws} in algorithms 
may make disconnected diagrams 
for multi-hadron calculations 
tractable in the not too distant future.
This means that 
pion-pion scattering phase shifts might be calculable soon within lattice QCD.
Actually, some first results for the isospin 2 waves
have been obtained for still somewhat large pion masses \cite{Sasaki:2008sv,Dudek:2010ew}.
Of course, lattice calculations still have systematic 
uncertainties which are hard to estimate 
and they always rely on modified actions, finite volumes, 
and other complications so that their physical results are actually
extrapolations to the physical limit. It is therefore 
necessary to understand how
these chiral or physical extrapolations should be carried out.

Fortunately, even though  we cannot rely on perturbative QCD at low energies, 
we can still use its effective low energy theory, 
known as Chiral Perturbation Theory
(ChPT) \cite{Gasser:1983yg},
which provides a rigorous, systematic and model independent expansion
of hadronic observables in terms of the external meson momenta 
and the relatively small pion mass. We will very briefly review ChPT in section~\ref{sec:ChPT}, mostly to introduce the required notation.

Within ChPT,  the quark mass dependence appears
in a model independent way through the pion mass squared, which is also described as an expansion.
Remarkably,
the isospin $I=2$  scattering length $m_\pi$ dependence found on the lattice
is rather well described by 
just leading order ChPT up to surprisingly
 large pion masses \cite{Beane:2007xs,Feng:2009ij} and the one-loop 
corrections seem to be rather small. In this work we will
first study the evolution of the lowest five 
pion-pion scattering phase shifts,
with definite isospin and angular momentum  
$(I,J)=(0,0),(1,1),(2,0),(0,2)$ and $(2,2)$,
 using the one and two-loop standard
ChPT expressions, estimating the uncertainties due to the relatively
poor knowledge of the low energy constants---particularly those at two loops. 
Of course, this approach is limited to low masses and momenta
and cannot be used to describe resonances, 
although, in principle it should be able to describe their low energy
tails, through, for instance, the low energy scattering phase shifts.
This is the reason why one of the aims of this work
is to study the evolution of  all 
$\pi\pi$ scattering phase shifts at low energy within standard ChPT.

Beyond the low energy regime, it is still possible to obtain 
the quark mass dependence of hadronic observables, by combining
ChPT with dispersion relations. 
Thus, in section \ref{sec:IAM} we briefly review
the Inverse Amplitude Method (IAM) 
\cite{Truong:1988zp,Dobado:1996ps,GomezNicola:2007qj}, obtained 
by using the
elastic approximation 
together with ChPT, to calculate the subtraction 
constants and the left cut contribution
of a dispersion relation for the inverse of the partial waves.
This technique provides a description 
of meson-meson scattering which is simultaneously 
compatible with the ChPT low energy description but  also generates
the lightest elastic resonance on each channel. 
By applying this technique
to the $\pi\pi$ scattering amplitude to one-loop in SU(2) ChPT, 
some of us have calculated the pion mass dependence of the
$\rho(770)$ and $f_0(600)$ masses and widths \cite{chiralexIAM}.
Interestingly, this method 
had already been applied to study only the $f_0(600)$
quark mass dependence and its influence, through the nucleon-nucleon interaction, on the production of carbon and oxygen and its anthropic implications
\cite{Donoghue:2006du}.
Recently \cite{Nebreda:2010wv}, some of us have also calculated 
the $\kappa(800)$ and $K^*(892)$ mass and width  
dependence  with respect to the non strange-quark mass, as well as the
dependence of all these four resonances with respect 
to the strange quark mass. And 
even more recently \cite{chiralexop6} we have extended 
to too loops 
the analysis of the $\rho(770)$ and $f_0(600)$ resonances within 
unitarized elastic $\pi\pi$ scattering.

The IAM results for the  $m_\pi$ dependence of the 
$\rho(770)$ agree nicely with the estimations 
for the two first coefficients of its chiral expansion \cite{bruns},
and also with the existing lattice results
\cite{lattice1,lattice2,lattice3,lattice4,lattice5}.
The comparison with lattice is relatively straightforward in this case since
the $\rho(770)$ is not extremely wide and 
it is actually calculated as a state of the spectrum.

Unfortunately, the comparison of the IAM  with lattice results
will not be so straightforward for the scalar channels. 
First, we find
of particular interest the repulsive $I=2$ channels.
Note that these channels have no resonances, so that neither the
 spectroscopic studies on the lattice
nor our  pole studies
 with the IAM \cite{chiralexIAM,Nebreda:2010wv,chiralexop6} address this case.
However, this is the simplest channel for scattering lattice 
studies and, as commented above,
there are already some lattice
results  for the scattering length down to relatively low pion masses
\cite{Beane:2007xs,Feng:2009ij}
and for  phase-shifts 
but only for $m_\pi\simeq 400$ MeV or higher \cite{Sasaki:2008sv,Dudek:2010ew}. 

Second, we are also interested in the much debated 
isoscalar channel.
Of course, given the status 
of the  $\sigma$ or  $f_0(600)$, reliable lattice
results would be most welcome. Unfortunately 
lattice calculations in this channel
are hard due to disconnected diagrams,
but also their interpretation would be complicated 
because this resonance is extremely wide (see \cite{PDG2010} and references therein).
In addition, it was shown in \cite{chiralexIAM,Nebreda:2010wv} that,
for sufficiently high masses, the $f_0(600)$,
being a scalar,  becomes a virtual state 
-- a pole in the second Riemann sheet below threshold -- 
which is not a physical state of the spectrum. 
Therefore, since spectroscopic (or ``pole'') lattice studies of the $\sigma$ may be rather complicated,
a study of the scalar phase shift, as the one presented here, deserves more interest.

These are the  motivations to study
the chiral extrapolation of phase shifts either from standard 
or unitarized ChPT. This will be done  first 
for standard ChPT to next to leading order (NLO)
in section~\ref{sec:resultsNLOChPT} and then to next 
to next to leading order (NNLO) in section~\ref{sec:resultsNNLOChPT}.
Surprisingly, in both cases,
the predicted behavior for the phase shift in the $\rho(770)$ 
may look counterintuitive when compared with present lattice calculations of the
$\rho(770)$ mass $m_\pi$ dependence. This discussion deserves a separated section, in which
we also evaluate the pion mass dependence of the ``size'' of the $\rho(770)$.
Next we will present the IAM results for 
NLO ChPT in section~\ref{sec:resultsNLOIAM} and for
NNLO in section~\ref{sec:resultsNNLOIAM}.
We will discuss and summarize all our findings in section~\ref{sec:summary}.

\section{Chiral Perturbation Theory}
\label{sec:ChPT}

Pions are the Goldstone bosons associated to the spontaneous 
chiral symmetry breaking of QCD. If quarks were strictly massless,
pions would be massless too and
separated by a gap of the order of 1 GeV from the rest of hadrons,
becoming the relevant QCD low energy degrees of freedom.
Chiral Perturbation Theory (ChPT) \cite{Gasser:1983yg} is nothing 
but the most general Lagrangian built out 
as an expansion in pion momenta (i.e., derivatives) respecting
the QCD symmetries. 
In real life, though, the $u$ and $d$ quarks have a very small mass,
that we will take in the isospin limit as $\hat m=(m_u+m_d)/2$,
which can be treated as a perturbation within ChPT.
As a consequence pions have a physical mass of $m_\pi=139.57\,$MeV, 
whose model independent perturbative expansion in terms of $\hat m$ is given
by ChPT.
In summary, the QCD low energy theory we will use is SU(2) ChPT \cite{Gasser:1983yg}, which corresponds
to considering the $u$ and $d$ quarks only and integrating out the other four quarks, whose
effect will be included in the low energy constants (LECs)
that multiply each term of the ChPT Lagrangian. In this way only pions will 
circulate in the loops. Hence, by varying the pion mass
while keeping the ChPT low energy constants fixed, we are sensitive to the light quark mass
dependence for constant $s,c,b$ and $t$ masses.

\subsection{Perturbative $\pi\pi$ scattering within ChPT}
\label{sec:perturbativepipi}

Pion-pion elastic scattering is customarily described in terms 
of partial wave
amplitudes $t_J^{(I)}(s)$ of definite isospin $I$ and angular momentum $J$, where $s$ is the Mandelstam variable,
although for simplicity we will drop these 
indices when there is no possible confusion. 
From ChPT these partial waves
 are obtained as a series expansion $t=t_2+t_4+t_6\cdots$, 
with $t_k=O(p/4\pi f_\pi)^k$,
where $p$ stands generically for center of mass momenta or pion masses.
The leading order (LO) $t_2$ is $O(p^2)$ and is universal \cite{chpt1} 
in the sense that it only depends 
on the scale $f_\pi\simeq92.4\,$MeV and $m_\pi$. The NLO calculation yields $t_4$ \cite{Gasser:1983yg} and
is obtained from one-loop diagrams
with LO vertices and tree diagrams from the NLO Lagrangian terms, which
are multiplied by some low energy constants (LECs), called $l_i^r(\mu)$.
These LECs absorb the dependence on the loop regularization scale $\mu$, 
and are determined by the underlying QCD dynamics.
Their measured values can be found in Table~\ref{standardLECS}.
Something similar happens with the NNLO result $t_6$ \cite{Bijnens:1997vq}, which has two loop
contributions with LO vertices, one-loop contributions with 
one LO vertex and one NLO vertex containing some $l_i$, 
plus tree level diagrams with NNLO vertices, whose LECs 
appear only in six combinations now called $r_i^r(\mu)$, whose estimated values are listed
also in Table~\ref{standardLECS}.
All these LECs carry a scale dependence that cancels that from loop integrals,
so that observables are scale independent and finite order by order.

\begin{table}
\begin{center}
    \begin{tabular}{cccc}
      \hline \hline
      \multicolumn{2}{c}{$O(p^{4})$ LECs ($\times 10^{-3}$)} &
         \multicolumn{2}{c}{$O(p^{6})$ LECs($\times 10^{-4}$)}\\
      \hline
            \noalign{\smallskip}
      $l_1^r$($\mu$) & -3.98 $\pm$  0.62 & $r_1^r$($\mu$) & -0.60 $\pm$ 0.35  \\
      $l_2^r$($\mu$) &  1.89 $\pm$ 0.23 & $r_2^r$($\mu$) & 1.28 $\pm$ 0.74     \\
      $l_3^r$($\mu$)   &  0.18 $\pm$ 1.11 & $r_3^r$($\mu$) & -1.68  $\pm$ 0.97\\
      $l_4^r$($\mu$) &  6.17 $\pm$ \ 1.39 & $r_4^r$($\mu$) & -1.00  $\pm$ 0.58 \\
      $ $&$ $ & $r_5^r$($\mu$) & 1.52 $\pm$ 0.42\\
      $ $&$ $ & $r_6^r$($\mu$) & 0.40 $\pm$ 0.04\\
            $ $&$ $ & $r_f^r$($\mu$) & 0.00 $\pm$ 1.20\\
      \hline
      \hline
    \end{tabular}
  \end{center}
  \caption{ ChPT low energy constants from \cite{Colangelo:2001df} that contribute
to $\pi\pi$ scattering to $O(p^4)$ and $O(p^6)$ that we use in our standard ChPT calculations. 
The value for $l_3^r$($\mu$) comes from a recent analysis of the lattice results
 \cite{Colangelo:2010et}.
The renormalization
scale is set to $\mu=770\,$MeV. Errors are only statistical
or ``only account for the noise seen in the calculations'' of \cite{Colangelo:2001df}.
The first four $r_i$ and their uncertainties are obtained from resonance saturation.
 The $r_f^r(\mu)$ value is from \cite{Colangelo:2003hf}. 
}
\label{standardLECS}
  \end{table}

Let us remark that we write the $\pi\pi$ scattering amplitude
in terms of the physical constants $m_\pi$ and $f_\pi$, which
are obtained as expansions in powers of the LO pion mass.
Actually, $l_3$ and $l_4$  appear at NLO in 
$\pi\pi$ scattering 
through these $m_\pi$ and $f_\pi$ expansions,
but in contributions 
that depend stronger on the pion mass and softer 
on the energy than those containng the other LECs. Thus $l_3$ and $l_4$ are harder to determine experimentally
and have the largest uncertainty.
This is particularly severe for $l_3$ and that is why 
we have used its lattice determination \cite{Colangelo:2010et}
quoted in Table~\ref{standardLECS}.

At NNLO, the expansion 
of $f_\pi$ on the physical pion mass \cite{Bijnens:2006zp}
requires an additional parameter $r_f$, also listed in Table~\ref{standardLECS}. 
Note that there is an additional $O(p^6)$  constant, $r_M$, 
 which appears in the NNLO chiral 
expansion of the physical pion mass $m_\pi$
in powers of the quark mass $\hat m$, but such a constant
would only be needed in order to study the quark 
mass dependence of observables. 
However, quark masses carry some renormalization scale and scheme
dependence and  most lattice results provide their results in 
terms of the physical pion mass. That is why here we will study the dependence of scattering phases
on the physical pion mass
and not on the quark mass. Therefore we do not need $r_M$.

We show in Table~\ref{standardLECS} the estimated statistical 
uncertainties of the LECs
(for $r_5,r_6$ they are described as the noise in
the dispersive calculation of~\cite{Colangelo:2001df}). 
Systematic uncertainties are large and harder to estimate;
for illustration we also provide in Table~\ref{tab:otherLECs}
other values found in the literature at $O(p^4)$ and $O(p^6)$.
We consider the spread on these values as a crude 
indication of the  size of systematic uncertainties.
From the sets in \cite{Amoros:2001cp} we note that, even for the same analysis, the values of the $O(p^4)$
LECs can be somewhat different whether
they are obtained from a pure $O(p^4)$ 
calculation or including the $O(p^6)$ corrections.
Hence, it should not come as a surprise later
that the $O(p^6)$ values obtained from a unitarized
fit, which includes part of the higher order corrections, may also come out somewhat different from the values obtained in a pure ChPT $O(p^6)$ analysis.

As a final comment concerning ChPT parameters, it is possible and usual to write 
the NNLO $\pi\pi$ scattering amplitude in terms 
of just six parameters $b_1, \ldots, b_6$, multiplying
each one of the energy dependent polynomials allowed by 
Lorentz invariance and chiral symmetry.
Thus, the knowledge of 6 constants 
is enough to describe $\pi\pi$ scattering
to that order.
However, these $b_i$ parameters do carry a dependence on $m_\pi$ 
and the full knowledge of all the $l_i$ and $r_i$ constants is needed
to extrapolate to unphysical values of $m_\pi$, which is the object of this work, and the reason why we need to determine eleven parameters instead of just six.

\begin{table}
  \label{LECsOp4}
  \begin{tabular}{|c|c|c|c|c|}
    \hline
    \vrule height 10pt depth 5pt width0pt
    Analysis & $10^3l_1^r$ & $10^3l_2^r$ & $10^3l_3^r$ & $10^3l_4^r$ \\
    \hline
ChPT    $O(p^4)$ \cite{Buettiker:2003pp} &$-4.9\pm0.6$&$5.2\pm0.1$& -- &$17\pm10$ \\
 ChPT    $O(p^4)$ \cite{Amoros:2001cp} &-4.5&5.9&2.1&5.7 \\
 ChPT    $O(p^6)$ \cite{Amoros:2001cp}&$-3.3\pm2.5$&$2.8\pm1.1$&$1.2\pm1.7$&$3.5\pm0.6$ \\
 ChPT    $O(p^6)$ \cite{Girlanda:1997ed} &$-4.0\pm2.1$&$1.6\pm1.0$&--&  --\\
\hline
IAM $O(p^4)$ \cite{chiralexIAM}&$-3.7\pm0.2$&$5.0\pm0.4$&$0.8\pm3.8$&$6.2\pm5.7$\\
    \hline
  \end{tabular}
  \caption{Samples of other sets of LECs:
First row:  SU(3) analysis of $\pi K$ scattering using Roy-Steiner equations.
Second  and third rows: $K_{l4}$ analysis to $O(p^4)$ and 
$O(p^6)$, respectively.  Naively, we have 
combined quadratically the SU(3) LECs errors there. 
Fourth row: Roy Equations analysis.
Uncertainties from imaginary parts and unknown 
$O(p^6)$ LECs combined quadratically.
Last row, values used in \cite{chiralexIAM} with the one-loop IAM.
    All LECs are evaluated at the scale $\mu=770$ MeV
}
\label{tab:otherLECs}
\end{table}

Now, elastic unitarity implies for
partial waves, at physical values of $s$, that:
\begin{equation}
  \label{elasticunit}
  \ima t(s)=\sigma (s) |t(s)|^2 \;\Rightarrow\; 
  \ima 1/t(s)=-\sigma (s),
\end{equation}
where $\sigma (s)=2p/\sqrt{s}$, 
$p$ being the center of mass momentum. As a consequence,
the modulus of $t(s)$ is related to its phase:
\begin{equation}
t(s)=|t(s)|e^{i\delta(s)}=e^{i\delta(s)}\sin\delta(s)/\sigma(s).
\end{equation}
This ``phase shift'' $\delta(s)$, which determines completely the 
amplitude, is the usual way to parametrize partial waves, that we will
use next to predict the amplitude variation when the pion mass is changed.
Of course, before extrapolating to other pion masses, 
we will compare the ChPT amplitudes, with and without unitarization,  
with the existing experimental  data.

ChPT amplitudes,
being an expansion, satisfy
unitarity only perturbatively:
\begin{equation}
  \label{unitpertu}
  \ima t_2=0,\quad\ima t_4=\sigma |t_2|^2,\quad
\ima t_6=2 \,t_2 \,\rea t_4\ldots.
\end{equation}
In particular, ChPT partial waves are expected to violate unitarity 
as $s$ increases, since they are basically polynomials in $s$.
In section~\ref{sec:IAM}  we will 
use ChPT inside dispersion relations to obtain amplitudes that,
while respecting the ChPT expansion at low energies, satisfy unitarity
and allow and provide a good description of experiment up to 
higher energies. 

After this brief introduction to ChPT and its notation, we are now ready to 
present our first calculations.

\section{Results within standard ChPT}
\label{sec:resultsChPT}

Using the equations above, the phase shift  within standard ChPT
is obtained as a series expansion
 (see \cite{Gasser:1990ku} for a prescription on how to perform this expansion):
\begin{eqnarray}
&&\delta=\sigma (t_2+\rea t_4)+O(p^6), \\
&&\delta=\sigma (t_2+\rea t_4+ \rea t_6)+ \frac{2}{3}(\sigma\, t_2)^3+O(p^8), \nonumber
\end{eqnarray}
which are the expressions used in our one-loop and two-loop
calculations, respectively, that we detail next.

Now, let us recall that the pion---and quark---mass dependence of the partial
waves $t(s)$  within ChPT comes from two different sources: from kinematics, through pion propagators, 
or from the dynamics encoded in the vertices. 
In particular, the threshold shift is purely of a kinematic 
nature and rather trivial to understand. Therefore, although $\pi\pi$ phase shifts are 
customarily presented in terms
of $\sqrt{s}$, we are showing them here as a function of the center of 
mass momentum $p$, which is also more convenient to compare to lattice studies.
With this kinematic threshold effect ``subtracted'', 
the remaining $m_\pi$ dependence is rather mild for most partial waves. 
 As we will see this soft dependence of the $\delta_{IJ}(p)$ 
on $m_\pi$ has been also found for $I=2$ waves in very recent lattice calculations \cite{Dudek:2010ew}.

\subsection{One-loop ChPT}
\label{sec:resultsNLOChPT}

\begin{figure}
  \includegraphics[scale=1.55]{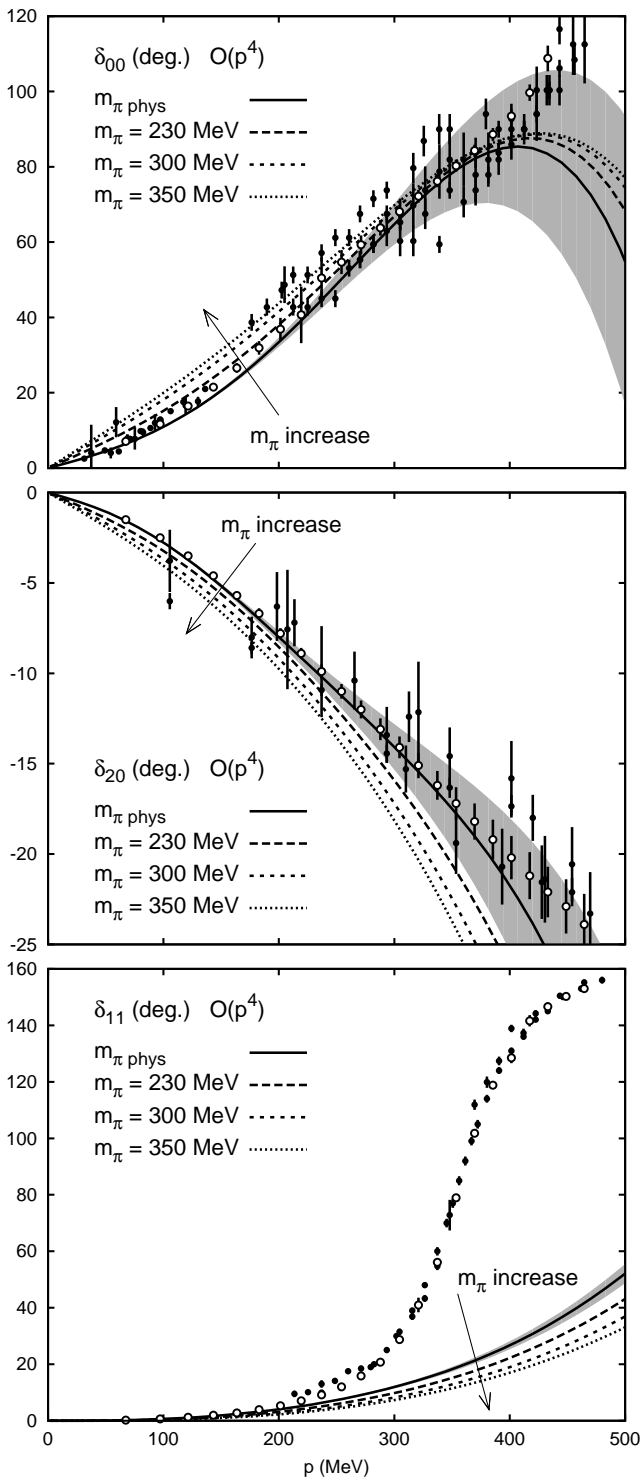}
    \caption{S and P wave $\pi \pi$ phase shifts from standard ChPT up to one loop. Different lines stand for different pion masses $m_\pi=139.57,\, 230,\, 300$ and 350~MeV, respectively. Since the lines are too close to each
 other, we only show error bands for the physical mass. 
Experimental data come from~\cite{experimentaldata}
(black circles) and the precise model independent 
dispersive data analysis from \cite{Martin:2011cn} (white circles).
The arrows show the direction of increasing $m_\pi$.
  \label{fig:NUOp4}}
\end{figure}

In Fig.~\ref{fig:NUOp4} we show the phase shifts from the one-loop ChPT, i.e. $O(p^4)$,
for the $(I,J)=(0,0), (2,0)$, $(1,1)$ $\pi\pi$ scattering waves. 
Note that for the $(1,1)$ channel, the description fails
much before $p\simeq 300$~MeV. 
This momentum is typically below the $\rho(770)$ resonance region, which
is a natural applicability bound for the ChPT series. 
This resonance has a relatively narrow shape, 
corresponding to a pole close to the real axis in the second Riemann sheet,
which, of course, is completely missed by one-loop ChPT  except
in its very low energy tail. In contrast, one-loop ChPT is giving a fairly good description
of the $(0,0)$ channel even up to, say $p=350$ or 400~MeV. In this case there is also a 
resonance---the scalar $\sigma$ (or $f_0(600)$)---, but it is very wide and its pole is deep
in the complex plane, so that it is not seen in the real axis as the typical sharp
rise in the phase. For this reason,
and despite being an expansion which has no such a pole in the complex plane,
 ChPT results are not very different qualitatively
from the data in this channel.
Finally, we see that 
the one-loop  description of the $(2,0)$ channel is also reasonably good
up to such high momentum, mostly due to the fact that this channel has no resonances
and also that the data are not particularly precise.

The gray areas in the figure cover the uncertainties 
due to the statistical error in the LECs detailed in the previous section.
In order to calculate these areas we have used a Montecarlo sampling. 
For each phase-shift calculation we have generated 5000 different 
samples of LECs using a Gaussian distribution with variances equal to 
the errors quoted in Table~\ref{standardLECS}. 
To avoid a confusing overlapping between uncertainty bands, we only show the one corresponding
to the physical pion mass. In the appendix we provide a detailed study of the evolution of these uncertainties with $m_\pi$. As a general feature
for both scalar and vector waves,
the relative uncertainty of the phase at a given momentum
grows slowly with  $m_\pi$.

Once we have checked where one-loop ChPT 
calculations provide an acceptable description of data, 
we can now compare, also in Fig.~\ref{fig:NUOp4}, with the 
the results obtained if we change the pion mass from its physical value 
to $m_\pi=230, 300$ and $350$~MeV. 
The first observation is that the sign of the phase derivative 
does not change when increasing the pion mass, at least up to 350~MeV, which means that 
the attractive or repulsive nature of each wave is conserved.

In that figure we have represented with an arrow 
the direction of the phase movement as $m_\pi$ increases.
Thus, the next observation is that both scalar phase shifts increase 
in absolute value as $m_\pi$ grows, whereas the phase of the vector channel decreases. 

The behavior of the phase at low momentum in the vector channel
may seem surprising at first, because several lattice works 
\cite{lattice1,lattice2,lattice3,lattice4,lattice5}, 
the chiral effective treatment \cite{bruns},
as well as the IAM \cite{chiralexIAM},
predict that the $\rho(770)$ mass increases much 
slower than the $2\pi$ threshold as $m_\pi$ grows. But then, when the $\rho(770)$ peak reaches a given
momentum, the phase there should be $\pi/2$ to a very good approximation. 
Therefore, one would expect naively the phase at low momentum to rise as $m_\pi$ grows.
However the model independent ChPT analysis, tells us otherwise.
We will see in detail in section \ref{sec:rhobehavior} why this intuitive picture fails and 
the phase shift actually has to decrease at first and increase later on.

\begin{figure}
  \includegraphics[scale=1.55]{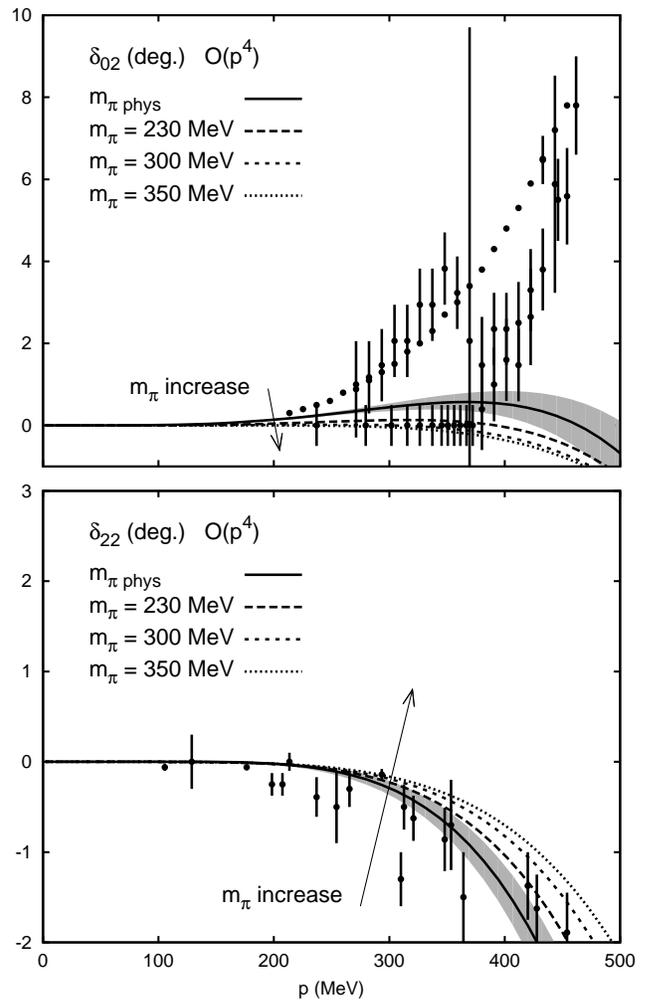}
    \caption{D wave $\pi \pi$ phase shifts from standard ChPT up to one loop. Different lines stand for different pion masses $m_\pi=139.57,\, 230,\, 300$ and 350~MeV, respectively. Since the lines are too close to each other, we only show error bands for the physical mass. Experimental data come from~\cite{experimentaldata}. 
The arrows show the direction of increasing $m_\pi$.
    }
  \label{fig:NUOp4D}
\end{figure}

Finally, in Fig.~\ref{fig:NUOp4D} we show the one-loop ChPT results 
for the D waves: $(I,J)=(0,2)$ and $(2,2)$.
We show these separately because both them vanish at $O(p^2)$, so that the one-loop $O(p^4)$ calculation is just their LO contribution. Actually they are both very small at low energies.
We can see in the figures that the  one-loop ChPT calculation provides an acceptable solution for the (2,2)
wave up to relatively high momentum, but obviously it cannot reproduce the resonance shape of the $f_2(1270)$ resonance in the $(0,2)$ channel. 
As before, we only show the uncertainty band due to the 
statistical errors on the LECs for the physical pion mass, obtained again from a Montecarlo Gaussian sample. Relative uncertainties for different pion masses are detailed in the appendix.

Note that, in contrast to the scalar waves, both tensor 
phase shifts decrease in absolute value as the pion mass increases not 
too far from its physical value. 
In this sense, they are more
similar to the vector channel behavior. Remarkably, for larger pion masses and momentum
the (0,2) phase shift even changes sign and 
the derivative becomes negative. However, this behavior is not found at two loops,
as we will see in the next subsection.

\subsection{Two-loop ChPT}
\label{sec:resultsNNLOChPT}

\begin{figure}
  \includegraphics[scale=1.55]{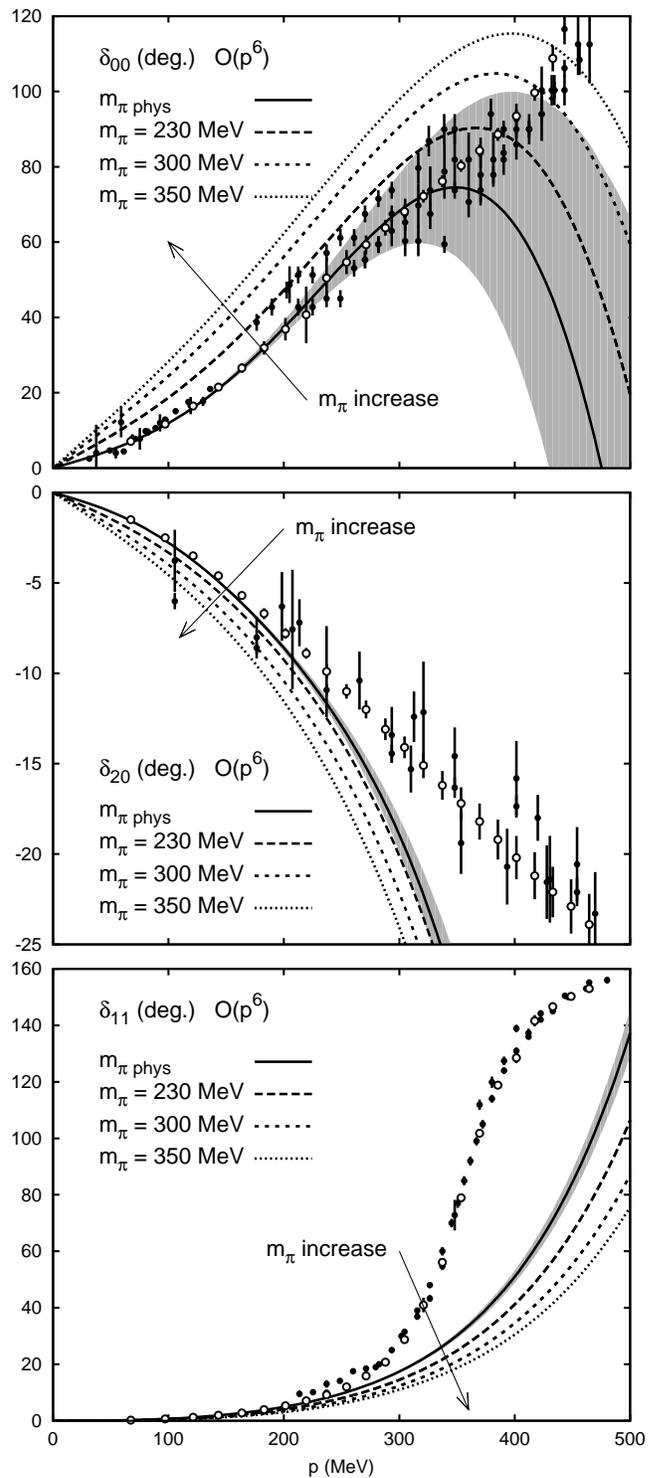}
    \caption{S and P wave $\pi \pi$ phase shifts from standard ChPT up to two loops. Different lines stand for different pion masses: continuous, long dashed, short dashed and dotted for $M_\pi=139.57,\, 230,\, 300$ and 350 MeV, respectively. Since the lines are too close to each other, we only show error bands for the physical mass. Experimental data (black circles) come from~\cite{experimentaldata} and the precise model independent 
dispersive data analysis from \cite{Martin:2011cn} (white circles). 
The arrows show the direction of increasing $m_\pi$.}
  \label{fig:NUOp6}
\end{figure}

We use the two-loop $\pi\pi$ scattering calculation in 
\cite{Bijnens:1997vq}. Note, however, that instead of the usual $b_1\ldots b_6$
parameters, in order to implement the $m_\pi$ 
dependence we need to use the one-loop $l_1\ldots l_4$ and $r_i$ parameters
in Table~\ref{standardLECS}. In Fig.~\ref{fig:NUOp6}
we show the resulting phase shifts 
for the $(I,J)=(0,0)$, (1,1) and (2,0) waves for the physical $m_\pi$ but also
for $m_\pi=230, 300$ and 350~MeV.

\begin{figure}
  \includegraphics[scale=1.55]{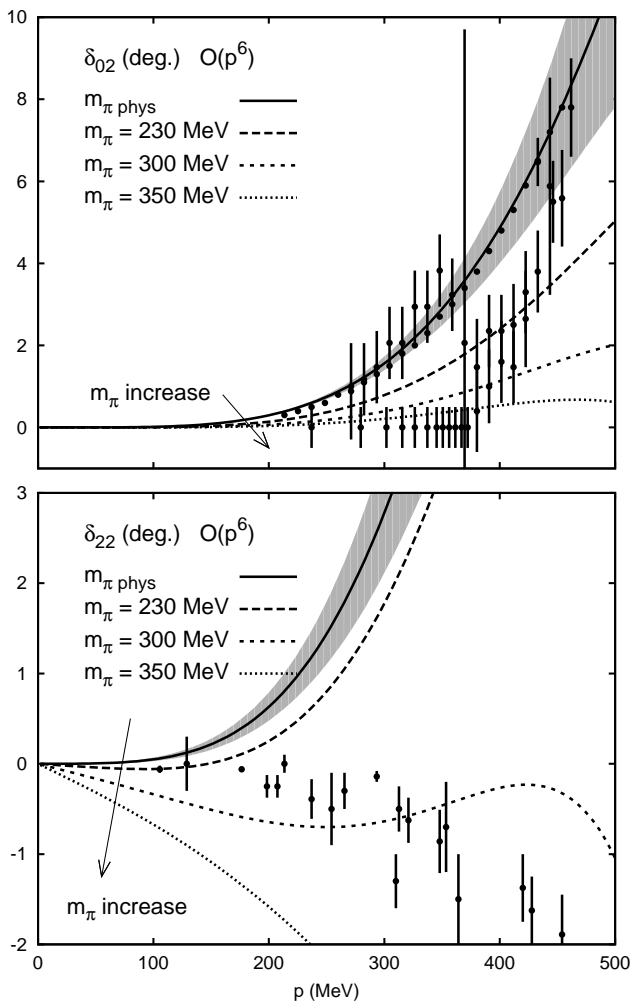}
 \caption{D wave $\pi \pi$ phase shifts from standard ChPT up to two loops. Different lines stand for different pion masses: continuous, long dashed, short dashed and dotted for $m_\pi=139.57,\, 230,\, 300$ and 350 MeV, respectively. Since the lines are too close to each other, we only show error bands for the physical mass. Experimental data (black circles) come from~\cite{experimentaldata}. 
The arrows show the direction of increasing $m_\pi$.}
  \label{fig:NUOp6D}
\end{figure}

The uncertainty bands, which we show only for the physical pion mass---see the appendix for other masses---are once again calculated with a Montecarlo Gaussian sampling of 5000 sets of
LECs, using as standard deviations the uncertainties quoted in Table~\ref{standardLECS}.
The only  exception are the $r_{1...4}$ parameters, which
are estimated from resonance saturation and, as in \cite{Colangelo:2001df}, 
we have assumed 
that all values in the interval from 0 to twice the estimation are equally 
likely. 
Of course, we want to emphasize that this is just an estimate of the values of the
$O(p^6)$ parameters, which are rather difficult to determine. Possible improvements
in their determinations could come from future lattice-QCD calculations, as it has already been
done with the $O(p^4)$ LECs (see \cite{Colangelo:2010et} for a review) or from the use
of recent dispesive data analysis like that in \cite{Martin:2011cn}
inside threshold parameter sum rules \cite{inprep}.

Also, since the renormalization scale $\mu$ 
where the estimates for $r_{1...4}$ and $r_f$ apply is not known, another source of uncertainty 
appears. Our calculations are made at  $\mu=770$ MeV so, in order to account 
for the uncertainty due to that choice, we have followed \cite{Colangelo:2001df} 
again and we have calculated the shift 
occurring in the phase shift if $r_{1...4}$ are fixed and the scale 
is changed to $\mu=500$~MeV and $\mu=1$~GeV. That shift is added 
in quadrature to the errors given by the Montecarlo sampling.

The general features of the one-loop description still 
apply to the two-loop case.
Namely, all waves keep their attractive or repulsive nature, and
both scalar phases increase in absolute value as $m_\pi$ grows, 
whereas the vector channel phase decreases.
The counterintuitive behavior of the $\rho(770)$ is therefore a robust
 prediction of ChPT. In the next section
we will explain with a simple model why chiral symmetry requires this behavior.
Still, the description of the (0,0) 
wave is fair only up to $p=300$ or 350 MeV, although
it has improved remarkably in the low energy region,
where the data are most recent and reliable, as they 
come from $K_{\ell 4}$ decays. 
The $(1,1)$ phase is now much closer to the experimental data, 
and thus it seems to provide a 
fairly good representation up to, say $p=200$~MeV.
However, the description of the $(2,0)$ has deteriorated for higher momenta, and
seems to be good only up to, roughly, 200 or 250~MeV.

However, despite the qualitative $m_\pi$  dependence being similar to the 
one-loop case, quantitatively the effect is stronger. In absolute value 
all phase shifts grow faster with $m_\pi$ to two loops than they did to
one loop.

In Fig.~\ref{fig:NUOp6D} we show the two-loop result for the D waves.
As commented before, these waves have no $O(p^2)$ term, so, this $O(p^6)$ 
calculation is just a next to leading order calculation. We can see that the differences with the one loop case are dramatic. The $(I,J)=(0,2)$
phase suffers a remarkable improvement, being able to describe the tail of the $f_2(1275)$ resonance up to momentum of the order of 350 MeV.
Contrary to the one-loop case, within  the $m_\pi$ range of this study, the $(0,2)$ phase does not
become negative.
Finally, the $(2,2)$ phase shift fails to describe even the 
sign of the data, and is only relatively close to the data points below 150~MeV.
Furthermore, the one-loop $m_\pi$ phase shift dependence was opposite  to 
the two-loop case: from more negative to less negative for the former
versus from positive to negative for the second. The predictions for this channel are therefore not very
robust, which is also corroborated by the large uncertainties for higher $m_\pi$ that can be found in the appendix.

\subsection{Comparison with lattice results for $I=2$ and $m_\pi>350$~MeV}

\begin{figure}
  \includegraphics[scale=1.5]{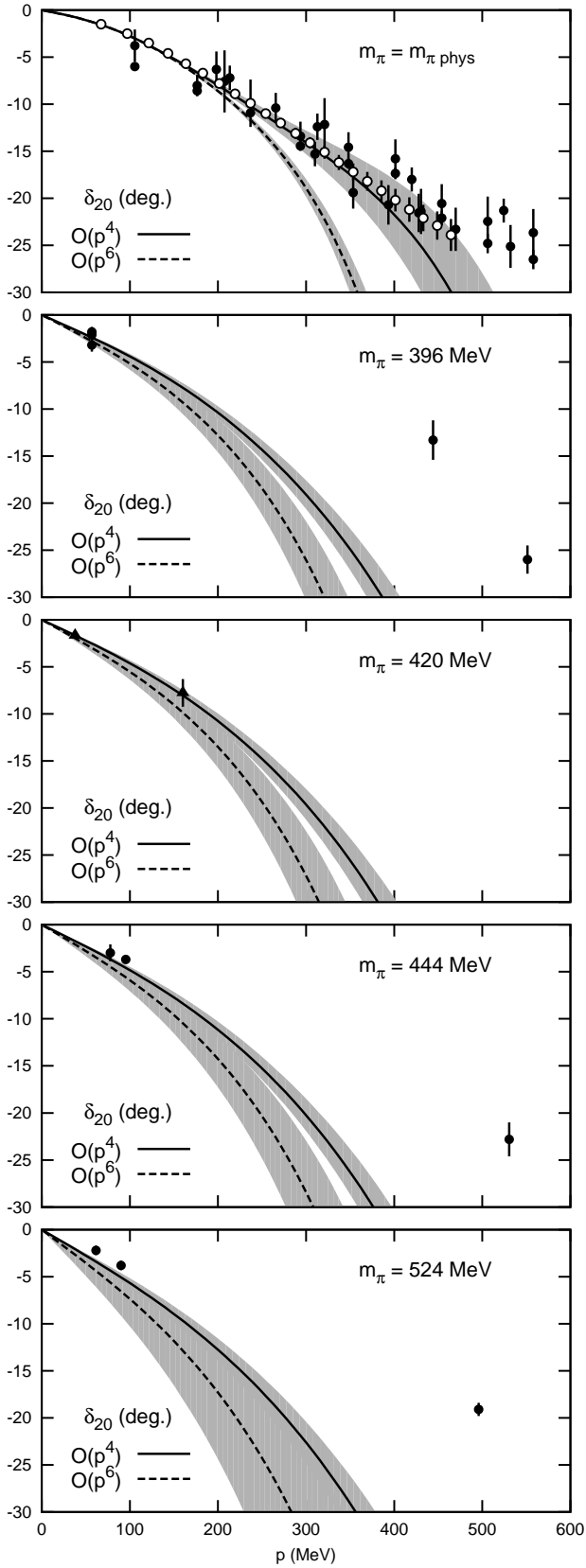}
    \caption{One and two loop standard ChPT phase shifts  for the $I=2$, $J=0$ channel compared to 
lattice results coming from~\cite{Dudek:2010ew} (circles) and~\cite{Sasaki:2008sv} (triangles).}
  \label{fig:lattice_NU_20}
\end{figure}

\begin{figure}
  \includegraphics[scale=1.5]{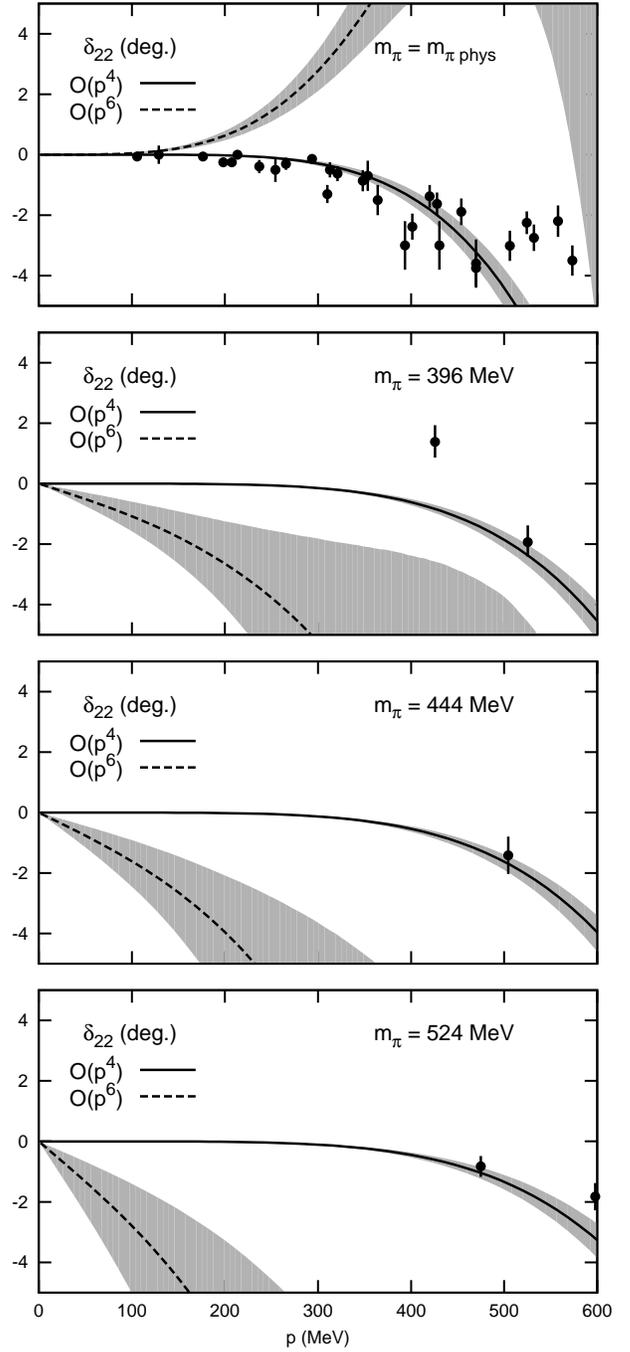}
    \caption{One and two loop standard ChPT phase shifts  for the $I=2$, $J=2$ channel compared to 
lattice results coming from~\cite{Dudek:2010ew} (circles). Note the large difference between one and two loop results.}
  \label{fig:lattice_NU_22}
\end{figure}

As we have already commented, there are 
very recent lattice results on phase shifts for the $I=2$, 
$J=0$ \cite{Sasaki:2008sv,Dudek:2010ew} and $J=2$ 
channels \cite{Dudek:2010ew}.
In Figs.~\ref{fig:lattice_NU_20} and \ref{fig:lattice_NU_22}  we compare
 the one and two-loop calculations within standard ChPT,
first for the physical mass versus experimental data, and then
for $m_\pi=396, 420, 444$ and 524~MeV, versus lattice results.

When we examine Fig.~\ref{fig:lattice_NU_20}, corresponding to  the $I=2$, 
$J=0$ phase shifts, the first observation is that all lattice points with $p<200$~MeV
are well described within the uncertainties of one-loop ChPT, even up to
$m_\pi=444$~ MeV. From the figure, we observe that
a pion mass of $524$~ MeV seems out of reach and will not be considered any longer.
Beyond that momentum, the ChPT calculation bends downwards and misses all other 
lattice results with higher momenta. 
Remarkably the two-loop ChPT results do not improve this agreement.
Actually, the two-loop calculation
describes somewhat worse the
lattice data and seems to move consistently 
to more negative values than those observed on the lattice, 
as $m_\pi$ grows higher. Let us remark that the curvature downwards
is larger in the two-loop result than just to one loop.
In view of the figures it seems that the standard ChPT applicability limit 
is, at best, somewhere around $p\simeq150-200$~MeV, up to $m_\pi$ of the order 
of $400-440$~MeV. 

Unfortunately, for the $I=2$, $J=2$ channel, shown in Fig.~\ref{fig:lattice_NU_22},  there are no
lattice results available at low momentum. 
Surprisingly, the one-loop calculation agrees quite 
nicely with the lattice values up to 
around $p\simeq 500$~MeV even for the highest pion mass. 
However, the two-loop results
show a very strong $m_\pi$ dependence that is in complete disagreement with the behavior predicted
by the lattice simulations. Even the tendency is wrong, since the
absolute value of the phase seems to grow with $m_\pi$ whereas lattice results may suggest a decrease. Let us nevertheless recall that for D-waves the tree level amplitude vanishes, so that one and two-loop calculations correspond only to leading and next to leading order results.
Higher order calculations may be needed to improve and stabilize the D wave description.

\section{Reconciling the phase shift and resonance behavior
in the vector channel}
\label{sec:rhobehavior}

We have seen that, within ChPT, the low momentum phase shift of the
vector channel is found to decrease as $m_\pi$ grows. 
This is a model independent result and 
looks rather robust since it is obtained both at one and two-loops.
However, lattice results \cite{lattice1,lattice2,lattice3,lattice4,lattice5},
the chiral effective treatment \cite{bruns},
as well as the IAM \cite{chiralexIAM}
predict that, in terms of momentum, the $\rho(770)$ peak gets 
closer and closer to threshold.
Thus, for any low momentum choice, and as $m_\pi$ increases, the $\rho(770)$ peak reaches 
that given momentum,
so that the phase there should be $\pi/2$. Therefore one would naively 
expect the phase shift for any fixed low momentum to grow with $m_\pi$.

Actually, this is what one would find if,
to describe the $\rho(770)$ resonance pole,
one uses the very simple and intuitive
(but, as we will see below, incomplete) Breit-Wigner model 
\begin{equation}
  \label{eq:T(s)}
  t(s)=\frac{-\sqrt{s} M\,\Gamma(p)/2p}{s-M^2+iM\Gamma(p)}
\end{equation}
where, $p^2=s/4-m_\pi^2$ and the width is:
\begin{equation}
  \Gamma(p)=\Gamma_R \left(\frac{p}{p_R}\right)^3,
\end{equation}
where $M$ is the resonance mass, 
$p_R$ is the pion momentum at the resonance energy $p_R^2=M^2/4-m_\pi^2$
so that $t(s)$ behaves correctly at threshold, $t(s)\sim p^{2l}$. 
Note that $\Gamma_R=g^2p_R^3/6\pi M^2$ is the $\rho(770)$ decay width.

For the sake of simplicity, let us now assume that the resonance mass $M$ 
and coupling 
remain constant when changing the pion mass $m_\pi$. 
This implies that $M$ and $\Gamma(p)$ are $m_\pi$ independent.
For our illustration purposes here, this is a fairly good approximation to what has been found
on the lattice or with the IAM, and it could be considered as the leading order
term in the $m_\pi$ expansion
(see  \cite{Bruns:2004tj} for the $\rho(770)$ mass). 

In such case, however, the phase-shift 
$m_\pi$ dependence near threshold does not follow what is obtained from
ChPT (or the IAM as we will see below). In particular, since
\begin{equation}
  \label{eq:tand}
  \tan\delta(p) = -\frac{M\Gamma(p)}{4p^2-4p_R^2},
\end{equation}
the only $m_\pi$ dependence in $\delta$ (for a given $p$) is through $p_R$
(and $d (p_R^2)/d(m_\pi^2)=-1$) so that
\begin{equation}
  \label{eq:ddelta/dm2}
  \frac{\partial\delta}{\partial(m_\pi^2)}=
  -\frac{\partial\delta}{\partial (p_R^2)}=
  \frac{4M\Gamma(p)}{\left(4p^2-4p_R^2\right)^2+M^2\Gamma(p)^2} >0.
\end{equation}
However, in ChPT, for low $p$ we have shown in 
Figs.~\ref{fig:NUOp4} and \ref{fig:NUOp6}
that $\partial\delta/\partial(m_\pi^2)<0$.

Of course, it is very well known that a simple Breit-Wigner vector formalism is not
consistent with the chiral expansion unless there are some additional low energy contributions
---or contact terms in the Lagrangian formalism \cite{Ecker:1989yg}. 
Just to keep things very simple we can use a modification of the 
Breit-Wigner parametrization, which is widely used in analysis of $\pi\pi$ scattering and other phenomenology involving decays into light mesons 
\cite{VonHippel:1972fg}, and reads
\begin{equation}
  \label{eq:GammaBlattWeiss}
  \Gamma(p)=\Gamma_{R}\left(\frac{p}{p_R}\right)^{2l+1}
  \frac{D_l(p_R r)}{D_l(pr)}\equiv 
  \tilde\Gamma(p)\frac{D_l(p_R r)}{D_l(pr)}.
\end{equation}
Here $\tilde\Gamma(p)$ is $m_\pi$ independent and $D_l(pr)$ are the Blatt-Weisskopf centrifugal barrier functions \cite{BlattWeisskopf}, 
that for $l=1$ read
$D_1(pr)=1+(pr)^2$. 
All the $m_\pi$ dependence is carried by $p_R$ and the new parameter $r$,
which is usually interpreted as a crude estimate of the ``size'' of the meson,
although it should not be identified with its mean square charge radius.
At low momentum we now find
\begin{equation}
  \label{eq:dps/dm2-low}
  \frac{\partial\delta(p)}{\partial(m_\pi^2)}\simeq
  \frac{1+p_R^4 (r^2)'}{4p_R^4}M\tilde\Gamma(p),
\end{equation}
where $(r^2)'$ stands for $dr^2/d(m_\pi^2)$. In order to have a
decreasing phase shift at low $p$ when increasing $m_\pi$,
we just need $1+p_R^4(r^2)'<0$. We will see below that this is actually 
required by chiral symmetry at leading order in the pion mass expansion.
This would explain the phase decrease seen in ChPT
for not too large $m_\pi$, even though the $\rho(770)$ is approaching threshold as $m_\pi$ grows. Of course,  when $m_\pi$ grows too large, and particularly
in the limit 
when the $\rho(770)$ 
tends to threshold, so that $p_R\rightarrow0$, the derivative is positive,
and the phase shift increases, as one would have
 expected naively.

Let us then check that chiral symmetry actually requires $1+p_R^4(r^2)'<0$ at least for
low pion masses.
We can estimate the leading $m_\pi$ dependence of $r$ by comparing the low
momentum and mass expansion of the amplitude in Eqs.\eqref{eq:T(s)}  using
\eqref{eq:GammaBlattWeiss},
with that of ChPT. 
In particular, since in this simple model we have only one parameter, 
$r$, we will only compare the scattering lengths.
Our aim is just to reproduce the 
leading order $m_\pi$ dependence, since 
we have already made
additional approximations and simplifications 
(like the constancy of the $\rho(770)$ mass and coupling).
We define the scattering length, 
$a$, as ${\rm Re}\, t\simeq p^2(a+bp^2+\cdots)$.
 The low $p$
expansion of the amplitude in Eqs.\eqref{eq:T(s)}  using
\eqref{eq:GammaBlattWeiss} leads to
\begin{eqnarray}
  \label{eq:aBW}
  a_{BW}&=&\frac{m_\pi M\Gamma_R\left(1+(p_Rr)^2\right)}{4p_R^5}\\
 & =&\frac{m_\pi\,\Gamma_R}{M\,p_R^3}\Big(1+\tfrac1{4}M^2r^2
  +O(m_\pi^2)\Big). \nonumber
\end{eqnarray}
This result has
to be compared with that of ChPT:
$a_{\rm ChPT}=1/24\pi f_\pi^2+O(m_\pi^2)$.
Matching with ChPT we obtain for $r^2$
\begin{eqnarray}
  \label{eq:r-chpt}
  r^2&=&\frac{p_R^3}{6\pi f_\pi^2 M\Gamma_R}\frac1{m_\pi}
  +O(m_\pi^0)\\
&\equiv& \frac{1}{g^2f_\pi^2}\frac{M}{m_\pi}+O(m_\pi^0)
  \simeq (4.3 \;{\rm GeV}^{-1})^2.
\end{eqnarray}
The value obtained with this ChPT estimation is compatible with
what is found in the literature ($r\sim 4-5$ GeV$^{-1})$) \cite{VonHippel:1972fg}. 

Note that the size $r$ explodes as $m_\pi\rightarrow0$. However,
this is a very well known feature of hadrons,
at least for the charge radius. Actually, the squared charged radius of the pion
and the nucleon show a $\log m_\pi^2$ singularity \cite{Gasser:1983yg,volkov,Beg:1973sc}
and the Pauli radius 
of the nucleon an additional $1/m_\pi$ singularity \cite{Beg:1973sc}.
Nevertheless,  as we have commented, our $r^2$ parameter
 should not be directly identified with the $\rho(770)$
charged radius, although our results suggest that they may have a similar singularity.

With this $m_\pi$ dependence for $r$ we find that
\begin{equation}
  1+p_R^4(r^2)'=1-\frac{M\,p_R^4}{2g^2 f_\pi^2m_\pi^3}
\end{equation}
which is negative for the physical values of the parameters. This guarantees that 
$\partial\delta(p)/\partial(m_\pi^2)<0$ for 
$m_\pi$ not far from $m_\pi^{phys}$, and sufficiently low $p$,
as is obtained in ChPT.  

The decrease is a robust feature of ChPT, 
although the pure chiral expansion cannot reproduce the $\rho(770)$ resonance.
Of course, the model we have presented here is very simple and naive, 
but provides a qualitative and intuitive explanation of why 
chiral symmetry implies that
the vector phase shift at low momenta first decreases, although it may increase later
 as $m_\pi$ grows. 
This model cannot be pushed too far. In particular, we cannot reproduce the 
chiral behavior of the scattering length beyond leading order or even the slope parameter.

It is however possible to incorporate simultaneously the $\rho(770)$ pole
and the full low energy ChPT expansion to one and two-loops. 
In the next section we will explain the technique in detail and later on we will show
how it describes the existing lattice data up to much higher momentum than standard ChPT.
Actually we will check how the vector phase shift 
decreases first and then increases as $m_\pi$ grows.

\section{Unitarized ChPT: The Inverse Amplitude Method}
\label{sec:IAM}

As we have already commented in Sect.~\ref{sec:perturbativepipi}, the
partial waves obtained from the ChPT expansion are basically a truncated series in momenta or energies and cannot satisfy elastic unitarity, Eq.\eqref{elasticunit}, exactly, but 
only perturbatively, as in Eq.\eqref{unitpertu}.

There is, however, a well known technique, known as unitarization,
 to obtain expressions for partial waves that satisfy elastic unitarity, have the correct analytic structure in terms of cuts in the complex plane, and simultaneously
respect the ChPT expansion up to a given order. Here we will make use of
the elastic Inverse Amplitude Method (IAM)---or a slightly modified version---that implements the 
fully renormalized one or two-loop ChPT 
expansion at low energies but does not introduce any spurious parameter in the unitarization procedure. Had we used other, possibly simpler but very successful,
unitarization techniques with spurious parameters, like cutoff or any other regulator,
we should have had to worry about the unknown $m_\pi$ dependence of that scale.

The IAM \cite{GomezNicola:2007qj} uses
elastic unitarity and the ChPT expansion to
evaluate a once subtracted dispersion relation
for the inverse amplitude. The analytic structure of
$1/t$ consists on a right cut from threshold to
$\infty$, a left cut from $-\infty$ to 0, and possible poles
coming from zeros of $t$. 
We can write then a once subtracted dispersion relation for
$1/t$, the subtraction point being $s_A$, 
\begin{equation}
  \begin{aligned}
    \label{disp1/t}
    \frac1{t(s)}&=
    \frac{s-s_A}{\pi}\int_{RC}ds'\frac{\ima 1/t(s')}{(s'-s_A)(s'-s)}\\
    &+LC(1/t)+PC(1/t),
  \end{aligned}
\end{equation}
where $LC(1/t)$ stands for a similar integral over the left cut
and $PC(1/t)$ is the contribution of the pole at $s_A$.
The choice of $s_A$ is, in principle, arbitrary, but since we want to use the 
information encoded in the ChPT series, 
we are then limited to the low energy region, preferably, below threshold.
Now, scalar waves vanish at the so called
Adler zero that lies in the real axis below threshold
and in practice this is a very convenient choice for $s_A$, which 
has actually motivated our notation. 
For other waves there is no such an Adler zero, and the subtraction
point can be taken, for instance at $s=0$. It is important to remark
that the choice of subtraction point, as long as it lies
 between the left and right cut, has only a very small numerical effect \cite{GomezNicola:2007qj} on the physical region.
Up to here everything is exact.
The most relevant observation is that, following Eq.~\eqref{elasticunit},
 on the elastic cut we know exactly
$\ima 1/t=-\sigma$.

Now we are going to derive the IAM within one-loop ChPT.
First, the Adler zero
position can be approximated as, $s_A=s_2+s_4+\cdots$,
where $t_2$ vanishes at $s_2$, $t_2+t_4$ vanishes at $s_2+s_4$, and so on.
On the right cut we can evaluate exactly
$\ima 1/t=-\sigma=-\ima t_4/t_2$, as can be read from Eqs. \eqref{elasticunit}
and \eqref{unitpertu}. Since the left cut is weighted at
low energies we can use one-loop 
ChPT to approximate $LC(1/t)\simeq LC(-t_4/t_2^2)$.
The pole contribution $PC(1/t)$ can be safely calculated with ChPT since
it involves derivatives of $t$ evaluated at $s_A$, which is a low energy
point where ChPT is perfectly justified. Altogether, we arrive to a
modified one-loop IAM (mIAM) formula \cite{GomezNicola:2007qj}:
\begin{equation}
  \begin{aligned}
    \label{mIAM}
    t^{mIAM}&=\frac{t_2^2}{t_2-t_4+A^{mIAM}},\\
    A^{mIAM}&=t_4(s_2)-\frac{(s_2-s_A)(s-s_2)[t'_2(s_2)-t'_4(s_2)]}{s-s_A},
  \end{aligned}
\end{equation}
where the prime denotes the first derivative with respect to $s$ and
where we use for $s_A$ in the numerical calculations its NLO 
approximation $s_2+s_4$.
The standard IAM formula is recovered for $A^{mIAM}=0$, which is indeed the
case for all partial waves except the scalar ones. 
In the original IAM
derivation \cite{Truong:1988zp,Dobado:1996ps} $A^{mIAM}$ was neglected
since it formally yields a higher order contribution and is numerically
very small except near the Adler zero. However, if $A^{mIAM}$ is
neglected, the IAM Adler zero occurs at $s_2$, correctly only to LO,
is a double zero instead of a simple one, and a spurious pole appears
close to the Adler zero. All of these caveats disappear with the mIAM, and
the differences between the IAM and the mIAM in the physical and resonance
region are of the order of 1\%.

It is important to remark that ChPT has {\it not} been used {\it at all}
for calculations of $t(s)$ for positive energies above threshold.
Note that the use of ChPT is well justified to calculate $s_A$, and $PC(1/t)$,
since these are low energy calculations. ChPT has also been used to calculate
the left cut integral, which, despite extending to infinity, 
is heavily weighted at low energies, which once again justifies the use of ChPT.
The left cut and the elastic approximation are the only approximations
used to obtain the IAM, but no other model dependent assumptions have been made.
In particular there are no spurious parameters
included in the IAM derivation, but just the ChPT LECs, $m_\pi$ and $f_\pi$.

Remarkably, these simple equations (either the IAM or the mIAM) 
ensure elastic unitarity, match ChPT
at low energies and,  using LECs 
compatible with existing determinations,
describe fairly well data up to somewhat 
less than 1 GeV, generating
the $\rho$, $K^ *$, $\sigma$ and $\kappa$ resonances
as poles on the
second Riemann sheet \cite{Dobado:1996ps}.

The extension to two loops is very similar and 
straightforward for the IAM \cite{Dobado:1996ps,Nieves:2001de}
or the mIAM \cite{chiralexop6}:
\begin{equation}
  \begin{aligned}
    t^{mIAM}&=\frac{t_2^2}{t_2-t_4+t_4^2/t_2-t_6+A^{mIAM}},\\
    A^{mIAM}&=t_4(s_2)-\frac{2t_4(s_2)t'_4(s_2)}{t'_2(s_2)}
    -\frac{t_4^2(s_2)}{t'_2(s_2)(s-s_2)}\\
    &+t_6(s_2)+\frac{(s-s_2)(s_A-s_2)}{s-s_A}
    \Bigg(
      t'_2(s_2)-t'_4(s_2)\Bigg.\\
      &\Bigg.-t'_6(s_2)+\frac{t'_4(s_2)^2+t''_4(s_2)t_4(s_2)}{t'_2(s_2)}
    \Bigg),\label{mIAMp6}
  \end{aligned}
\end{equation}

Let us now remark that both in the one and two-loop derivations above, we have assumed that $t_2$ is not identically zero. However, this is only the case for scalar and vector partial waves. Unfortunately, as seen in Eq.\eqref{unitpertu},
when $t_2(s)\equiv 0$ the first imaginary part appears at $O(p^8)$, namely, at three loops. Therefore we cannot recast the dispersion relation in terms of
the full ChPT expansion unless we make use of $t_8(s)$, a 
calculation that does not exist. In \cite{Dobado:2001rv}, and using only the $t_8$
term of the form  $c s^4$, it was shown that the $f_2(1275)$ shape could be
fairly well fitted with the IAM and a $c$ value of the 
correct order of magnitude expected from dimensional grounds.
This was justified 
because the $f_2$ resonance appears at  high $s>>m_\pi^2$ and the other $O(p^8)$
terms, containing pion mass powers, could be neglected.
However, in this work we want to make $m_\pi$ much
 larger than its physical value and we need the $m_\pi$ dependence.
It is therefore not so well justified to neglect all the $t_8$ terms except $cs^4$.
For that reason we are limited to use the IAM for scalar and vector partial waves.

Hence, using the IAM or the mIAM,  we can study how the 
generated $\rho$ and $\sigma$ poles evolve by changing $m_\pi$ in the one loop IAM amplitudes
\cite{chiralexIAM} or two-loop amplitudes \cite{chiralexop6}, and describe
the dependence of their masses, widths and
couplings on $m_\pi$.
In \cite{chiralexIAM}
the mIAM was used for the $\rho$ and $\sigma$ chiral extrapolation,
because, for the scalar and  at high $m_\pi$,
one resonance pole gets near the IAM spurious pole,
a problem that is nicely solved with the mIAM. Nevertheless,
in the physical region and near the other generated poles, the
differences between IAM and mIAM approaches are almost negligible, even for
high pion masses. 

Of course, the poles are not the only object of study on the lattice.
Actually, lattice results are already available 
for phase shifts in $I=2$ channels, where no pole exists.
Moreover these channels were not studied
in \cite{chiralexIAM,chiralexop6}.
It is also very likely that lattice results on phase shifts
 for other channels will be available soon.
For these reasons we will now let $m_\pi$ vary within our unitarized 
ChPT expressions, with the aim of extending the phase shift predictions 
based on ChPT, up to higher masses and momenta.

\section{Results with the IAM and ChPT}
\label{sec:resultsIAM}

Let us first recall, as already explained in some of the very first works on the IAM \cite{Dobado:1996ps},
and repeated in many other instances \cite{GomezNicola:2001as,Pelaez:2006qv,chiralexop6},
that when the central values of the standard LECs are used, the IAM only improves 
ChPT up to a couple of hundred MeV higher and resonances are 
only reproduced qualitatively. For a semi-quantitative
 description of resonances, which is what we will do next, one has to fit the data and the resulting LECs are slightly modified
from those obtained from pure ChPT. 
Since the IAM contains contributions that count as higher order in ChPT (in particular the numerically relevant s-channel logarithms), one would very naively expect the LECs 
from the one-loop IAM to lie somewhere in between the one and two loop values from ChPT.  
This is actually observed, since the $O(p^4)$ IAM LECs in Table~\ref{tab:LECsIAMNLO} lie
somewhere between the one and two loop analysis of pure ChPT listed in Table~\ref{tab:otherLECs},
although closer to the ChPT $O(p^4)$ analysis in the two first rows of that table. In contrast, the $O(p^4)$ values of the LECs for the two-loop IAM in Table~\ref{tab:LECsIAMNNLO} are closer to the 
two loop analyses like that in Table~\ref{standardLECS} or those in the third and fourth row of Table~\ref{tab:otherLECs}. Let us emphasize that the variation between the $O(p^4)$ LECs values between
the one and two loop analyses already occurs in pure ChPT---particularly for $l_2^r$. The IAM simply follows a similar pattern.

Before changing the pion mass, let us note that for the IAM
 we are assuming the elastic approximation and 
therefore, when increasing $m_\pi$ we should allow for 
some $\pi\pi$ elastic regime, which is guaranteed if
$m_\pi< 500$ MeV, although it has been found that relatively
stable unitarized results can be obtained for all waves only 
up to $m_\pi\simeq300-350\,$MeV \cite{chiralexop6}.
Of course some waves are more stable 
than others. In particular the elastic IAM
approximation is quite good up to larger 
energies 
for the $(I,J)=(2,0)$ (roughly up to $\sqrt{s}\simeq 1200-1300$, see \cite{GomezNicola:2001as}), 
since
it has no resonances and does not couple to $\bar KK$.
We will actually check that for this 
channel we can stretch the applicability range and
still get fairly good agreement with recent 
lattice results for relatively large pion masses.

\subsection{One-loop IAM}
\label{sec:resultsNLOIAM}

\begin{figure}
  \includegraphics[scale=1.55]{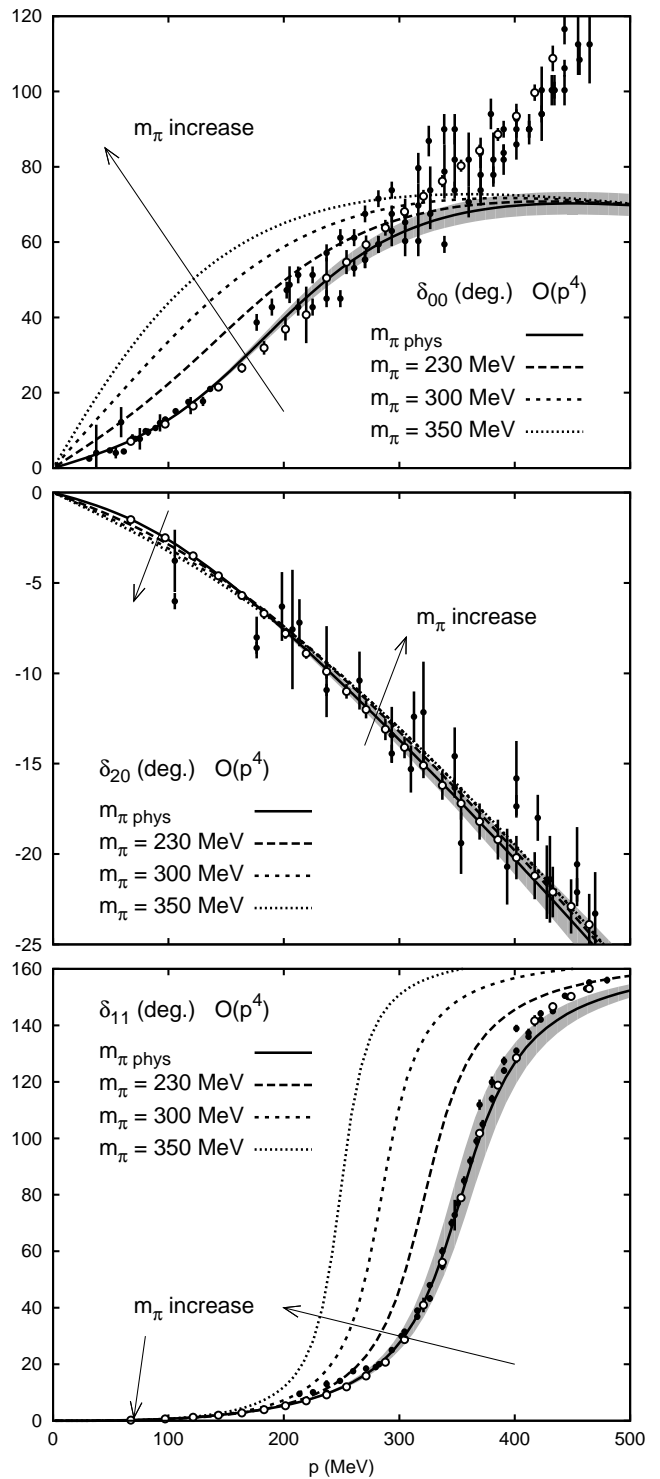}
    \caption{S and P wave $\pi \pi$ phase shifts from unitarized ChPT up to one loop. Different lines stand for different pion masses: continuous, long dashed, short dashed and dotted for $M_\pi=139.57,\, 230,\, 300$ and 350 MeV, respectively. Since the lines are too close to each other, we only show error bands for the physical mass.
Experimental data come from~\cite{experimentaldata}
(black circles) and the precise model independent 
dispersive data analysis from \cite{Martin:2011cn} (white circles).
The arrows show the direction of increasing $m_\pi$. See Fig.~\ref{fig:zoom11} for a blow up of the low momentum region of the $I=1$, $J=1$ phase shift.}
  \label{fig:Uphaseshifts}
\end{figure}

In Fig.~\ref{fig:Uphaseshifts} we show the IAM results to one-loop in ChPT,
using the LECs in Table~\ref{tab:LECsIAMNLO}, obtained by an updated
fit to
the output from the recent and precise dispersive data analysis
in \cite{Martin:2011cn}, and fixing $l_3$ 
and $l_4$ to the updated values in Table~\ref{standardLECS}.  
The uncertainties are mostly systematic, arising
from different choices of the maximum energy up to where we make the fit of the $(0,0)$ channel, which we have chosen between 500 and 800 MeV, the other channels are
fitted up to 1 GeV. Note that the resulting LECs are consistent
within one standard deviation with the results we used in \cite{chiralexIAM},
that we list in the last row of Table~\ref{tab:otherLECs}.
We first note that the
experimental data is fairly well described up to the region 
where inelastic effects (or resonances like the $f_0(980)$) become relevant. 
This includes the $\rho(770)$ resonance shape, but also the wide shape of the $f_0(600)$. 
The gray bands in the figures cover the uncertainties in our results
obtained from a Monte Carlo Gaussian sampling of 
the $l_i$ statistical error bars also listed in the table.
As usual, and to avoid confusion due to many overlapping gray bands,
we only show the uncertainty for the physical pion mass. 
Details on uncertainties for higher masses can be found in the appendix.

\begin{table}
  \begin{center}
    \begin{tabular}{cc}
      \hline
      \hline
 \multicolumn{2}{c}{$O(p^{4})$ LECs ($\times 10^{-3}$)}\\     
     \hline
      $l_1^r$($\mu$) & $-3.9 \pm 0.2$  \\
      $l_2^r $($\mu$) &  $4.3 \pm 0.4$ \\
      $l_3^r $($\mu$)   & $0.18 \pm 1.11$\\
      $l_4^r $($\mu$) & $6.17 \pm 1.39$\\
      \hline
      \hline
    \end{tabular}
  \end{center}
  \caption{ LECs used in this work for the one-loop IAM, obtained 
from a fit to the dispersive data analysis of \cite{Martin:2011cn}.
Both $l_3$ and $l_4$ are fixed to the standard values given in Table~\ref{standardLECS}. 
The scale is set to $\mu=770$~MeV.}
  \label{tab:LECsIAMNLO}
\end{table}

The general features for the scalar isoscalar channel are very similar to the 
one-loop non-unitarized results. Namely, the phase shift
conserve its positive sign and increases in absolute value as $m_\pi$ grows.

However, the $I=2$ channel behavior is rather different.
First, the $m_\pi$ dependence is even milder than for the non-unitarized case.
In the very low momentum region, roughly below $p=200$~MeV,
the phase increases in absolute value as it happened with standard one-loop
ChPT. However, for larger momentum, the $m_\pi$ 
dependence is the opposite, and the phase starts decreasing its absolute value.
As we will see later on, this is the behavior found on recent lattice
results, which cannot be reproduced by a crude extrapolation of one-loop
ChPT to larger momentum.

\begin{figure}
  \includegraphics[scale=1.55]{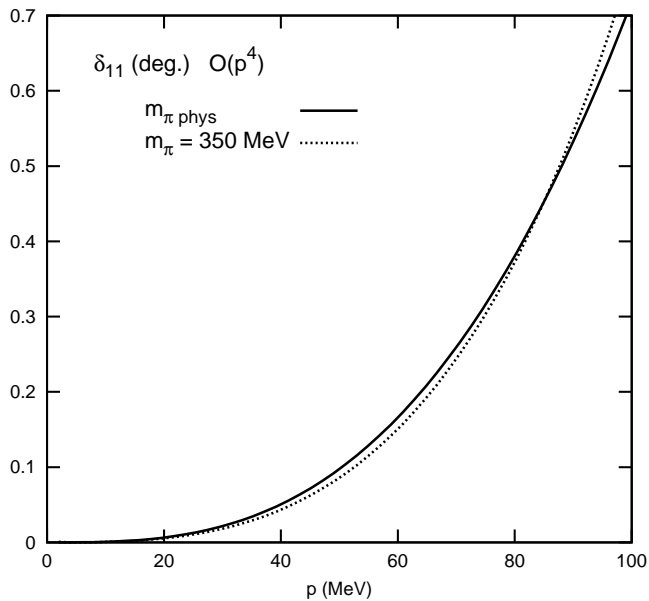}
    \caption{$\pi \pi$ $I=1$, $J=1$ phase shift from unitarized ChPT up to one loop. The continuous line stands for $M_\pi=139.57$~ MeV and the dotted line for $M_\pi=350$ MeV. Similarly to the ChPT case, in the low momentum region the phase shift decreases as $m_\pi$ grows.
However at higher momentum it increases with the pion mass, due
to the presence of the $\rho(770)$ resonance. }
  \label{fig:zoom11}
\end{figure}

Something similar occurs in the vector channel, 
although  enhanced by the presence of the 
$\rho(770)$ resonance that ChPT failed to reproduce.
Now we see that the phase increases as the two pion threshold
grows and gets closer to the resonance. This is the intuitive
behavior one would expect when getting close to the resonance.
However, one should observe that it is {\it not} incompatible with the 
phase decrease observed in standard ChPT at low energies. 
To see this, in Fig.~\ref{fig:zoom11}, we show a blow 
up of the very low energy 
region of the vector channel, 
where we can see that the IAM behaves similarly to ChPT, namely, the 
phase decreases as $m_\pi$ grows. 
As explained before, this only happens
in the very low momentum regime,
since, as seen in the figure, for higher momentum
the phase shift increases again since
the IAM is able to reconstruct the $\rho(770)$ resonance,
which is closer and closer to threshold as $m_\pi$ grows.

In the next subsection we will see that these general features 
and improvements with respect to non-unitarized ChPT are even more 
dramatic when considering the two-loop calculation.

\subsection{Two-loop IAM}
\label{sec:resultsNNLOIAM}

\begin{figure}
  \includegraphics[scale=1.55]{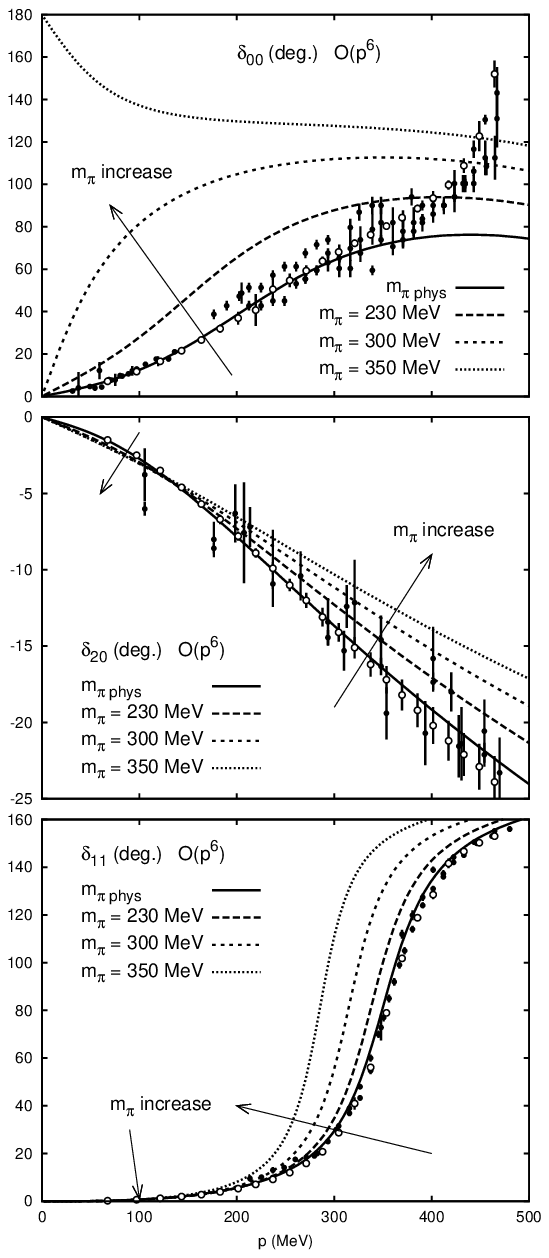}
    \caption{S and P wave $\pi \pi$ phase shifts from the two-loop IAM ``fit A'' in \cite{chiralexop6}. 
The conventions are as in Fig.~\ref{fig:Uphaseshifts}.
The arrows show the direction of increasing $m_\pi$. The difference between these curves and those
in Fig.\ref{fig:IAMtwoloopsD} are an indication of the order of magnitude of our uncertainties.}
  \label{fig:IAMtwoloopsA}
\end{figure}

\begin{figure}
  \includegraphics[scale=1.55]{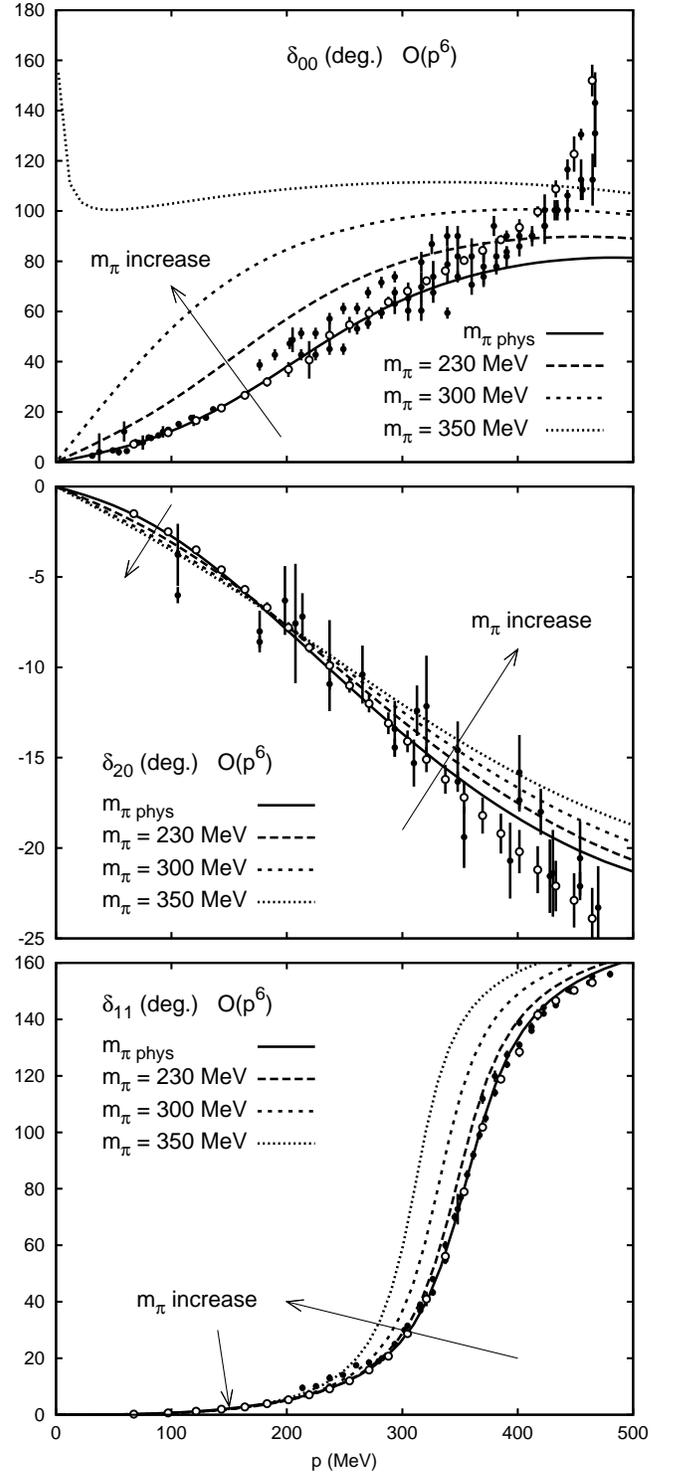}
    \caption{S and P wave $\pi \pi$ phase shifts from the two-loop IAM ``fit D'' in \cite{chiralexop6}. 
The conventions are as in Fig.~\ref{fig:Uphaseshifts}. 
The arrows show the direction of increasing $m_\pi$. The difference between these curves and those
in Fig.\ref{fig:IAMtwoloopsA} are an indication of the order of magnitude of our uncertainties.}
  \label{fig:IAMtwoloopsD}
\end{figure}

In Figs.~\ref{fig:IAMtwoloopsA} and \ref{fig:IAMtwoloopsD} we show the results of the two-loop IAM
for the two best fits in \cite{chiralexop6}, ``A'' and ``D'', 
whose corresponding sets of LECs we provide in Table~\ref{tab:LECsIAMNNLO}.
These fits have been obtained from an IAM fit to experimental data but also to lattice results
on $f_\pi$, $M_\rho$ and the isospin 2 scattering length. Note that by fitting only the 
experimental data one determines better the LECs that govern the $s$ dependence, but
not so well those governing the $m_\pi$ dependence. That is the reason why some existing lattice results on $f_\pi$, $M_\rho$ and the $I=2$ scalar scattering length were also
included in the fits of \cite{chiralexop6}. Unfortunately, the experimental data in the resonance region
are frequently in conflict with one another, and to a lesser extent, something similar happens for 
the lattice results mentioned above. Fits A and D correspond to different ways of 
weighting the conflicting  experimental and lattice results, including some educated estimates for systematic uncertainties. The details can be found in \cite{chiralexop6}. These fits give rather stable
results for all observables in the elastic region, up to $m_\pi=300-350$~MeV, and somewhat beyond
 for some particular waves, like $(I,J)=(2,0)$.

\begin{table}
\begin{center}
    \begin{tabular}{ccc}
      \hline \hline
      & Set A & Set D \\
      \hline
      \noalign{\smallskip}
      $O(p^{4})(\text{x}10^{-3})$ & & \\
      $l_1^r$($\mu$) &   -5.0 & -4.0 \\
      $l_2^r$($\mu$) &  1.7 & 1.2 \\
      $l_3^r$($\mu$)   & 0.8 & 0.8\\
      $l_4^r$($\mu$) & 6.5 & 6.5 \\
            \noalign{\smallskip}
      \hline
            \noalign{\smallskip}
      $O(p^{6})(\text{x}10^{-4})$ & & \\
$r_1^r$($\mu$) & -0.6 & -0.6 \\
$r_2^r$($\mu$) & 1.3 & 1.5 \\
$r_3^r$($\mu$) & -1.7 & -3.3 \\
$r_4^r$($\mu$) & 2.0 & 0.9 \\
$r_5^r$($\mu$) & 2.0 & 1.7 \\
$r_6^r$($\mu$) & -0.6 & -0.7 \\
$r_f^r$($\mu$) & -1.4 & -1.8 \\
      \noalign{\smallskip}
      \hline
      \hline
    \end{tabular}
  \end{center}
\caption{Low energy constants obtained from a fits \cite{chiralexop6}
to experimental data on elastic $\pi\pi$ scattering
and lattice results on $f_\pi$, $M_\rho$ and the isospin 2 scattering length as well as a $1/N_c$ leading 
behavior of a pure $\bar qq$ state.
Many of these sets are not quite compatible with each other and suffer large systematic uncertainties. 
These two fits correspond to different ways of weighting the existing experimental and lattice data sets,
which are detailed in   \cite{chiralexop6}. The values correspond to the scale $\mu=770$~MeV.}
\label{tab:LECsIAMNNLO}
\end{table}

Note that the qualitative behavior of all waves is similar
in Figs.~\ref{fig:IAMtwoloopsA} and \ref{fig:IAMtwoloopsD}. 
The difference between fit A and D is purely quantitative: in fit A the 
$m_\pi$ dependence is just stronger than in fit D. 

Remarkably, almost all the features described for the one-loop unitarized case remain in the two-loop unitarized fits.
Quantitatively there are small differences, since
 the $m_\pi$ dependence at two loops seems somewhat stronger  in the scalar waves, and somewhat weaker in the vector channel.
 This somewhat stronger $m_\pi$ dependence produces the only significant, and relevant, difference with the one-loop IAM.
Both the one and two-loop IAM generate the $f_0(600)$ or $\sigma$ resonance as a pole deep in the complex plane, 
which mass grows much slower than the two pion threshold, so that the "`bump"' that this wide resonance
produces in the $(0,0)$ phase is bigger and gets closer to threshold. Actually, as shown in \cite{chiralexIAM}
the two conjugated poles of the 
$f_0(600)$  move in the second, unphysical, Riemann sheet, until they reach the real axis below threshold, where the two poles
are no longer conjugated. As $m_\pi$ keeps growing 
one of them jumps into the first Riemann sheet below threshold becoming a bound state.
By Levinson's theorem \cite{Levinson}, this implies that the phase at threshold increases by $\pi$.
For the IAM to one-loop this jump occurs for $m_\pi$ larger than 350~MeV, but since the $m_\pi$ dependence 
is stronger for the IAM at two loops, this jump can already be seen in Figs.~\ref{fig:IAMtwoloopsA} and \ref{fig:IAMtwoloopsD}
for the $m_\pi=350$~MeV, which behavior thus reflects the existence of a bound state.
Let us emphasize that the same behavior would be observed to one loop---although for higher $m_\pi$---but it will
never be seen in standard ChPT, which cannot generate a pole.

However, when comparing with the non-unitarized two-loop results in Fig.~\ref{fig:NUOp6}, 
we see that unitarization not only
improves the vector channel by describing the $\rho(770)$ resonance, but also the $I=2$ channel
is nicely
described up to much higher momentum, even though this channel is non-resonant.
We will profit from this lack of complicated resonant structures in the $I=2$ scalar wave,
and also from the fact that this channel does not couple to $\bar KK$,
 to extrapolate to higher pion masses where we will see that the unitarized results are 
in much better agreement than standard ChPT with some recent lattice results.

\section{Comparison with lattice results for $I=2$ and $m_\pi>350$~MeV}

\begin{figure}[h]
  \includegraphics[scale=1.5]{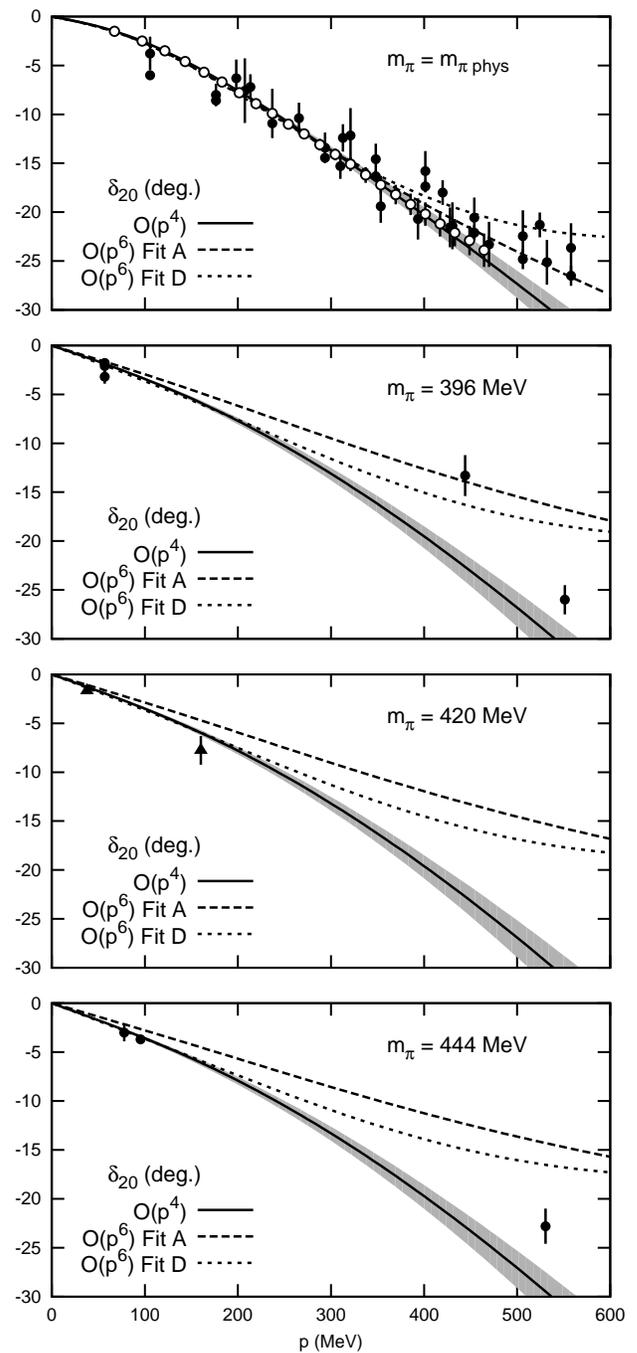}
    \caption{One and two-loop IAM phase shifts for the $I=2$, $J=0$ channel compared to 
lattice results coming from ~\cite{Dudek:2010ew} (circles) and ~\cite{Sasaki:2008sv} (triangles). Note that for the two-loop case we provide results for the two best fits, A and D, obtained in \cite{chiralexop6}.}
  \label{fig:lattice_U}
\end{figure}

In Fig.\ref{fig:lattice_U} we show the results from the one and two-loop IAM with 
very recent results on the lattice \cite{Sasaki:2008sv,Dudek:2010ew} for the $I=2$ scalar channel.
Note that the data below $p=200$~MeV is still fairly well described by the IAM, as it happened with ChPT, but that the IAM is not bending down and getting away from higher momentum data as it happened with standard ChPT results. Actually, the IAM results follow qualitatively the shape of the 
lattice data. Moreover, the $m_\pi$ dependence is much milder than for plain ChPT, in better agreement with the findings on the lattice. Let us remark that we do not aim at precision here because pion masses of 400 MeV are probably 
close to the IAM applicability bound. 
Our approach should become more reliable below 300-350~MeV, where we expect lattice results to appear soon. Still,  the remarkable
improvement with respect to the standard ChPT results is pretty clear.

As previously commented, the IAM cannot be directly applied to the D waves, since their tree level 
contribution vanishes. Further modifications of the IAM would be needed which are
beyond the scope of this work.

\section{Summary and discussion}
\label{sec:summary}

In this work we have studied the pion mass dependence of $\pi\pi$ elastic scattering phase shifts.

On the one hand we have presented results for one and two-loop standard Chiral Perturbation Theory  using a set of LECs obtained from a dispersive analysis
in the literature. We have seen that this first approach is, of course, limited to 
low momentum, say below 300~MeV, depending on the channel,
and pion masses up to 400-450 MeV. 
For the scalar and vector waves, we have found a rather stable behavior between the one and two
loop calculations within that momentum range.
We have seen that at this very low momentum, 
the absolute value of scalar phase shifts increases as the pion mass grows, 
so that these channels enhance their attractive or repulsive nature.
We have found that up to momenta less than 200~MeV, the ChPT results are in fair agreement with lattice data for the scalar $I=2$ channel.

We have found that, surprisingly, the vector phase-shift at very low momentum
 decreases as $m_\pi$ grows within the applicability region. This may seem counterintuitive,
since from lattice and other effective theory techniques, as $m_\pi$ grows one expects the two-pion threshold
to approach fast the $\rho(770)$ mass. We have nevertheless shown with a very simple and intuitive 
model why very basic requirements about chiral symmetry impose such a decrease on the phase
for low momentum and not too large $m_\pi$.

We have also shown results within standard ChPT for the angular momentum 2 phase shifts.
These are much less stable when comparing one and two loop results. 
Particularly for the $(I,J)=(2,2)$
channel, the one and two loop results show an opposite behavior, and the two loop calculation
is also at odds with the $m_\pi$ dependence found on the lattice.
Of course, one has to keep in mind that for D waves the one and two-loop calculations
correspond to leading and next to leading order calculations, 
contrary to scalar and vector channels, where they correspond to next to leading and next to next to leading calculations. It is very likely that higher order calculations, or better determinations of LECs, which are highly correlated, may improve this situation for D waves.

Finally, we have used ChPT inside a dispersion relation to extend the
analysis of scalar and vector waves to higher momentum by means of the so called Inverse Amplitude Method.
This unitarization technique describes remarkably well the data up to energies of the 
order of 1 or 1.2~GeV, depending on the channel and has been shown to describe well
the $m_\pi$ dependence of several observables like $M_\rho$, $f_\pi$ or the $I=2$ 
scalar scattering length.

The description provided by this method is of course compatible with that of standard ChPT
at very low momentum. However at higher momentum it reconstructs the behavior of the $\rho(770)$
resonance, which, for a given choice of low momentum, translates into a decreasing phase for smaller $m_\pi$ 
but a growing phase for larger $m_\pi$ until the $\rho(770)$ mass coincides with that particular 
momentum choice. In addition, we have shown that the unitarized $I=2$ scalar phase shift
has the correct qualitative behavior for momentum beyond 200~MeV. Despite being close to
the applicability bounds of the approach, we have actually shown that the IAM beyond $p=150$-200~MeV
improves dramatically the description of lattice results with respect to ChPT
and explains their very mild $m_\pi$ dependence.

Intuitively, the phase shift evolution of the S0 and P channels is dominated by the presence
of the $f_0(600)$ and $\rho(770)$ resonances and their pion mass dependence, studied in 
\cite{bruns,chiralexIAM,Nebreda:2010wv,chiralexop6}. Since the masses of both resonances 
seem to grow slower than the pion mass, they come closer and closer to threshold, so that, 
naively one would expect the interaction to grow stronger and 
the phase to raise once the resonance is sufficiently close to the  momentum
where the phase is measured. Actually, this is what is found for the S0 channel, whose phase raises noticeably
as $m_\pi$ grows.  At the limit of the range of applicability of the two-loop IAM, the $f_0(600)$ 
even becomes a bound state and by Levinson's theorem we see 
the phase to increase by $\pi$ at threshold. However, the naive expectations may
 not be met 
if the resonance is still not close enough to threshold. In such case, the phase may
seem to decrease at first due to the finite size of the resonance, which effect
has been illustrated
in a simple model of the $\rho(770)$. Only when the $\rho(770)$ is sufficiently close to threshold, 
the naively expected behavior is observed. Concerning the S2 wave, we have found a very mild $m_\pi$ dependence
for the phase shift, when expressed in terms of the momentum, in good agreement
with recent lattice calculations. This can be understood from the absence of resonant structures in this channel. Of course, ChPT can only reproduce the low energy tails of the resonances, which we have generated by means of 
ChPT unitarized with the IAM. For the D waves, the IAM cannot be applied to this order,
and we have to rely on ChPT only. However,
the behavior observed can also be understood from the presence of the $f_2(1270)$ resonance in the D0
channel, and a similar behavior to the $\rho(770)$ in its own channel. For the D2 channel the ChPT results
are not sufficiently precise to make any conclusive statement.

Apart from understanding the dependence of these observables on QCD parameters 
on the pion mass,
we consider that this work is of interest as a guideline for future studies of lattice QCD.

\section*{Acknowledgments}
We thank J. Dudek for lattice results and detailed explanations.
Work partially supported by Spanish Ministerio de 
Educaci\'on y Ciencia research contracts: FPA2007-29115-E,
FPA2008-00592 and FIS2006-03438, 
U.Complutense/Banco Santander grant PR34/07-15875-BSCH and
UCM-BSCH GR58/08 910309. We acknowledge the support 
of the European Community-Research Infrastructure
Integrating Activity
“Study of Strongly Interacting Matter” 
(acronym HadronPhysics2, Grant Agreement
n. 227431)
under the Seventh Framework Programme of EU.

\appendix
\section{Phase shift uncertainties for different $m_\pi$}

\begin{figure*}
  \includegraphics[scale=1.5]{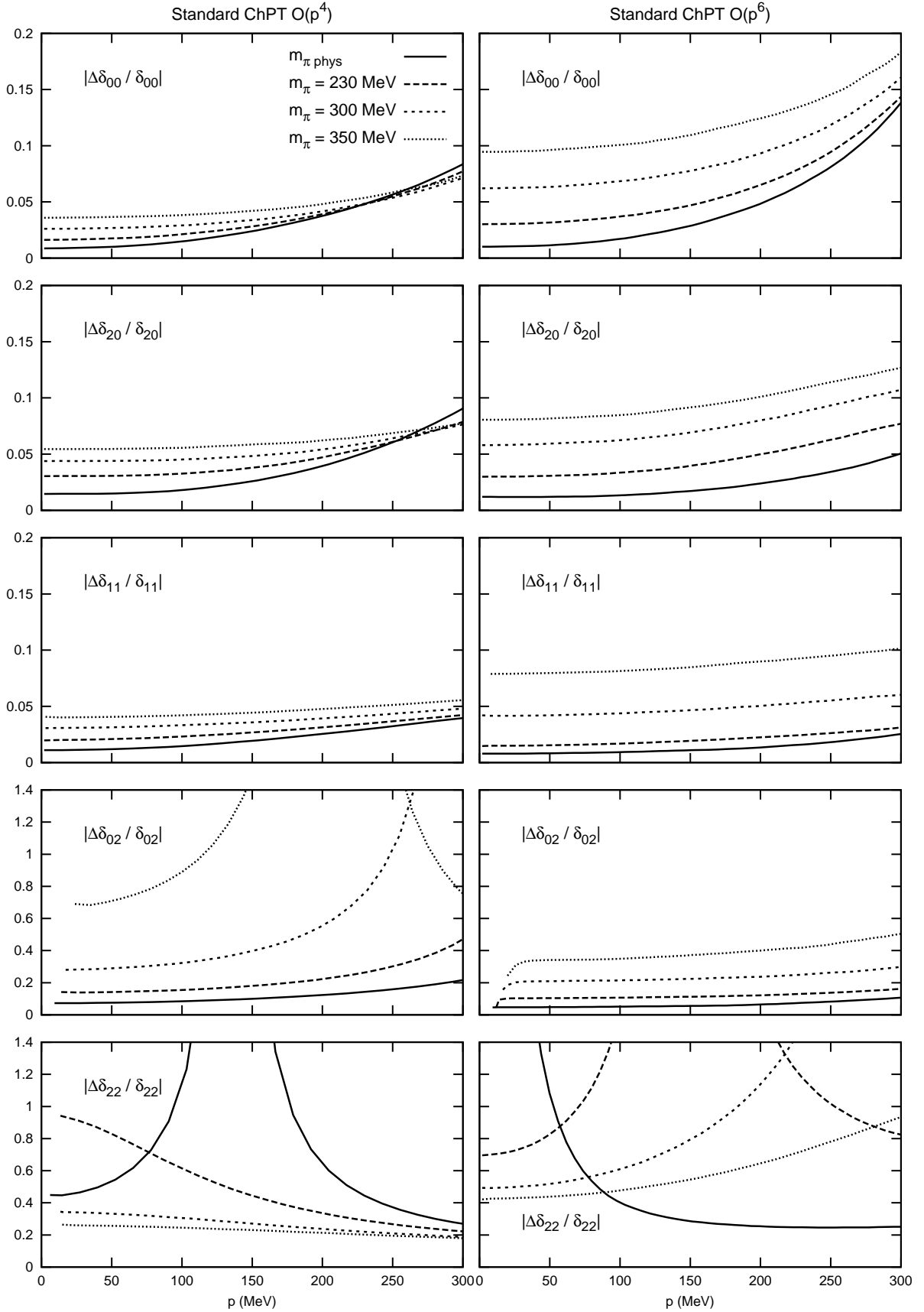}
    \caption{$\pi\pi$ phase shift errors normalized to the value of the phase shifts in standard ChPT to one loop (left column) and to two loops (right column). Different lines stand for different pion masses: continuous, long dashed, short dashed and dotted for 
$m_\pi=139.57,\, 230,\, 300$ and 350 MeV, respectively. }
  \label{fig:errores_StChPT}
\end{figure*}

In Fig.~\ref{fig:errores_StChPT} we plot the relative uncertainties
 of the standard ChPT phase shift calculation.  
As we have already seen, standard ChPT is limited to low momentum and thus 
we only show momentum up to $p = 300$ MeV. For the scalar and vector waves we see 
that in the low momentum region the errors grow with the pion mass. This is in 
agreement with the fact that the LECs that govern the mass dependence of the partial 
waves carry the biggest uncertainties. 
For D-waves, the relative uncertainty is much bigger than for lower angular 
momentum waves. (Note the difference in scales between the D waves and the rest of the plots). This is due to the fact that for D-waves the tree level calculation vanishes and therefore the one and two loop results are just leading and next to leading order.
In the case of $\delta_{02}$ to one loop 
the error seems to explode for the highest masses due to the phase shift 
changing from a positive to a negative value in the region of interest. The same occurs
for $\delta_{22}$ to one loop for the physical value of the pion mass. Finally, the value of $\delta_{22}$ to two loops changes from negative to positive for the lightest masses of the pion.

\begin{figure*}
  \includegraphics[scale=1.4]{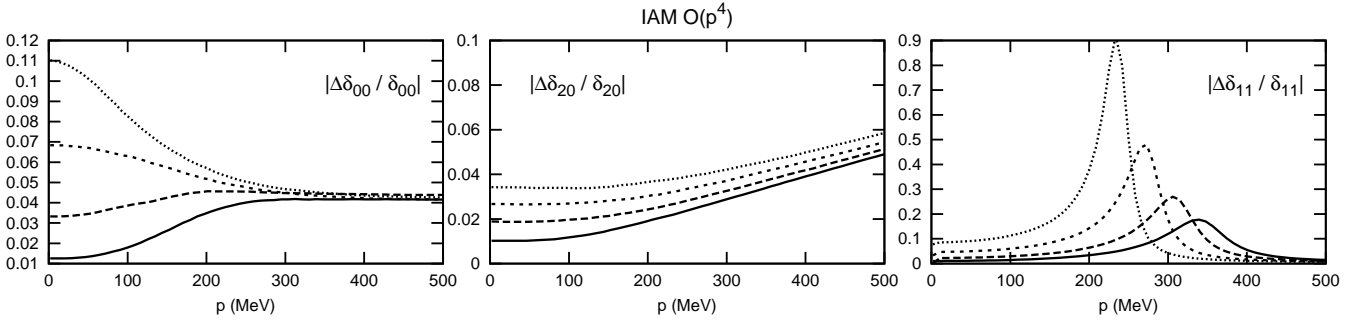}
    \caption{$\pi\pi$ phase shift errors normalized to the value of the phase shifts in unitarized ChPT to one loop. Different lines stand for different pion masses: continuous, long dashed, short dashed and dotted for $m_\pi=139.57,\, 230,\, 300$ and 350 MeV, respectively.}
  \label{fig:errores_IAM}
\end{figure*}

In Fig.~\ref{fig:errores_IAM} we show the relative uncertainties
 for the IAM phase shifts. We find again that for scalar waves 
they grow bigger as the pion mass is increased. 
The same happens for the vector phase shift below the $\rho(770)$ peak.
The highest uncertainty on $\delta_{11}$ 
occurs when the slope of the phase shift  reaches its maximum value. 

\end{document}